\documentclass[aps,twocolumn,showpacs,superscriptaddress,preprintnumbers,longbibliography]{revtex4-2} 
\usepackage{graphicx}  
\usepackage{dcolumn}   
\usepackage{bm}        
\usepackage{amssymb}   
\usepackage{amsmath}
\usepackage{comment}
\usepackage{graphicx}
\usepackage{color}
\usepackage{slashed}
\usepackage{braket}
\usepackage{amssymb,amsmath,graphicx,color}
\usepackage[tight]{subfigure}
\usepackage[export]{adjustbox}
\usepackage[colorlinks=true,citecolor=blue,linkcolor=red]{hyperref}
\usepackage{multirow}
\usepackage{rotating,booktabs}
\usepackage[verbose]{placeins}
\usepackage{mathrsfs}
\usepackage{tikz}
\usetikzlibrary{quantikz2}
\usepackage{algorithm}
\usepackage{algpseudocode}
\usepackage{soul}
\usepackage{lineno}

\newcommand{\Eq}[1]{Eq.~(\ref{#1})}

\newcommand{\Fig}[1]{Fig.~\ref{#1}}
\newcommand{\Figure}[1]{Figure~\ref{#1}}

\newcommand{\ua}{\mathord{\uparrow}}
\newcommand{\da}{\mathord{\downarrow}}

\begin{document}


\title{
Quantum Computing Universal Thermalization Dynamics 
\\
in a (2+1)D Lattice Gauge Theory 
}

\author{Niklas Mueller}
\email{niklasmu@unm.edu}
\affiliation{Center for Quantum Information and Control, University of New Mexico, Albuquerque, NM 87106, USA}
\affiliation{Department of Physics and Astronomy, University of New Mexico, Albuquerque, NM 87106, USA} 
\affiliation{InQubator for Quantum Simulation (IQuS), Department of Physics,
University of Washington, Seattle, WA 98195, USA}

\author{Tianyi Wang}
\affiliation{Department of Physics, Duke University, Durham, NC 27708, USA}
\affiliation{Duke Quantum Center, Duke University, Durham, NC 27701, USA}
\affiliation{The NSF Institute for Robust Quantum Simulation, University of Maryland, College Park, Maryland 20742, USA}
\author{Or Katz}
\affiliation{Duke Quantum Center, Duke University, Durham, NC 27701, USA}
\affiliation{Department of Electrical and Computer Engineering, Duke University, Durham, NC 27708, USA}
\affiliation{School of Applied and Engineering Physics, Cornell University, Ithaca, NY 14853, USA}
\author{Zohreh Davoudi}
\affiliation{Department of Physics and Maryland Center for Fundamental Physics, University of Maryland, College Park, MD 20742 USA}
\affiliation{Joint Center for Quantum Information and Computer Science, NIST and University of Maryland, College Park, MD 20742 USA}
\affiliation{The NSF Institute for Robust Quantum Simulation, University of Maryland, College Park, Maryland 20742, USA}
\affiliation{National Quantum Laboratory (QLab), University of Maryland, College Park, MD 20742 USA}
\author{Marko Cetina}
\affiliation{Department of Physics, Duke University, Durham, NC 27708, USA}
\affiliation{Duke Quantum Center, Duke University, Durham, NC 27701, USA}
\affiliation{Department of Electrical and Computer Engineering, Duke University, Durham, NC 27708, USA}
\affiliation{The NSF Institute for Robust Quantum Simulation, University of Maryland, College Park, Maryland 20742, USA}


\begin{abstract}
Simulating non-equilibrium phenomena in strongly-interacting quantum many-body systems, including thermalization, is a promising application of near-term and future quantum computation.
By performing experiments on a digital quantum computer consisting of fully-connected optically-controlled trapped ions, we study the role of entanglement in the thermalization dynamics of a $Z_2$ lattice gauge theory in 2+1 spacetime dimensions. 
Using randomized-measurement protocols, we efficiently learn a classical approximation of non-equilibrium states that yields the gap-ratio distribution and the spectral form factor of the entanglement Hamiltonian. 
These observables exhibit universal early-time signals for quantum chaos, a prerequisite for thermalization.
Our work, therefore, establishes quantum computers as robust tools for studying universal features of thermalization in complex many-body systems, including in gauge theories.
\end{abstract}

\maketitle

\section{Introduction}
\noindent
Thermalization of isolated quantum many-body systems
in e.g., ultra-cold atomic gases~\cite{eisert2015quantum,schreiber2015observation,schachenmayer2015thermalization,kaufman2016quantum,eigen2018universal,ueda2020quantum, yuan2023ooh,zhou2022thermalization}, trapped ions \cite{smith2016mbl, neyenhuis2017oop, morong2021oos, kyprianidis2021oop}, condensed matter physics~\cite{nandkishore2015many,borgonovi2016quantum}, cosmology~\cite{micha2004turbulent}, and nuclear and high-energy physics~\cite{baier2001bottom,berges2021qcd}, remains a vibrant frontier.  
Most quantum systems thermalize according to the Eigenstate Thermalization Hypothesis (ETH)~\cite{deutsch1991quantum,srednicki1994chaos}, which
posits that, under broadly applicable conditions, the long-time average of certain observables agrees with microcanonical predictions~\cite{rigol2008thermalization,d2016quantum}. Yet, probing large quantum many-body systems {\it undergoing} thermalization is inherently challenging due to the non-equilibrium nature of the process, which precludes using first-principle classical computational techniques. Recent advancements in quantum information theory and experiment~\cite{grumbling2019quantum,arute2019quantum,deutsch2020harnessing,bluvstein2021controlling,evered2023high} have brought this topic within immediate experimental reach, potentially enabling, via quantum simulation~\cite{cirac2012goals,daley2022practical,bauer2023quantum}, verification of important thermalization paradigms, and discovery of novel principles grounded in quantum information science.

While it is commonly posited that quantum chaos and ergodicity are prerequisites for thermalization~\cite{haake1991quantum,goldstein2010normal}, their demonstration in the context of quantum many-body systems remains somewhat elusive. Indicators of chaos and ergodicity involve measures associated with the eigenvalues of a given Hamiltonian~\cite{berry1977level,bohigas1984characterization,joshi2022probing}, or the properties of its eigenstates~\cite{pandey2020adiabatic}. A connection has been drawn recently between 
the {entanglement} of quantum states and quantum chaos, via the so-called entanglement Hamiltonians (EHs)~\cite{li2008entanglement,dalmonte2022entanglement}. This connection can, in principle, be leveraged to experimentally probe thermalization via quantum simulation. Most studies to date, nonetheless, have stayed in the realm of theoretical exploration~\cite{oganesyan2007localization,rakovszky2019signatures,chang2019evolution,mueller2022thermalization,ma2022quantum,froland2024nongaussian}.

Analog quantum simulation allows one to monitor a quantum system continuously in time. However, current analog quantum simulators have limited programmability and are restricted to probing specific physical models and a limited set of observables~\cite{smith2016mbl, neyenhuis2017oop, morong2021oos, kyprianidis2021oop}. In contrast, universal digital quantum computers allow, in principle, the probing of the dynamics of {any} physical model. The universal control in digital quantum computers enables the use of tomographic techniques to extract a wide range of observables, including entanglement.  
Digitized time evolution, via Trotterization~\cite{suzuki1991general,childs2021theory} or other schemes~\cite{lin2022lecture,childs2017lecture}, limits the near-term simulations to times that are short compared to those of thermalization. Nonetheless, there exist universal phenomena that are indicative of quantum chaos and ergodicity at earlier times. Probing these phenomena, therefore, is an attractive near-term opportunity for digital quantum computers.
\begin{figure*}[t]
  \centering
        \includegraphics[width=0.99\textwidth]{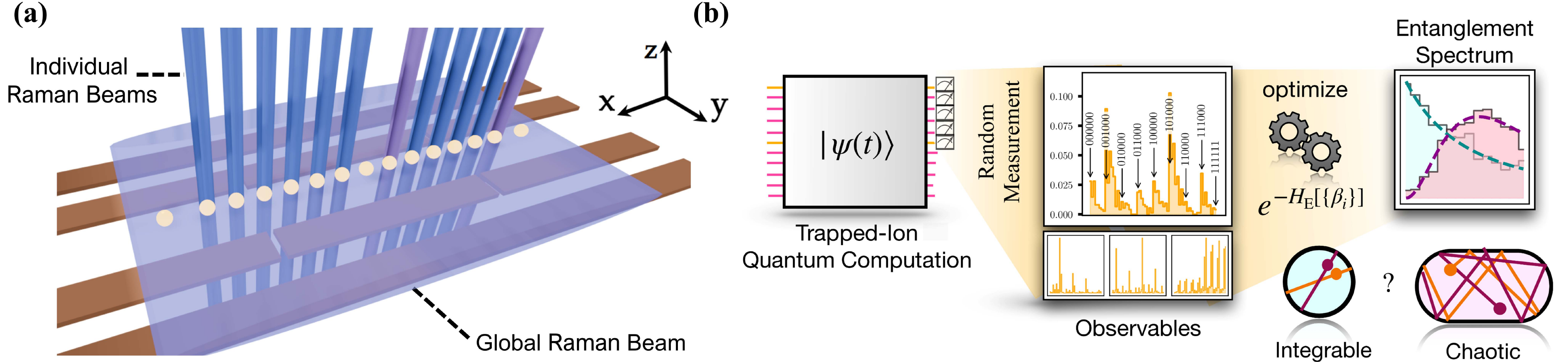}
	\caption{ \emph{Overview.} (a) Schematic of the trapped-ion experiment: 15 optically-controlled ions {(yellow circles)} in a linear trap {(electrodes shown as brown rectangular pads)} realize a universal digital quantum computer. Single-qubit and all-to-all two-qubit gates are implemented by an array of individually-focused laser beams {(blue and purple, vertical)} and a global laser beam (blue, horizontal). Two-qubit gates are performed using pairs of individual beams, such as the ones highlighted in purple. (b) Schematic of the randomized-measurement protocol for entanglement-Hamiltonian tomography. This protocol extracts a classical approximation of a reduced density matrix associated with a subsystem of the quantum state $\ket{\psi(t)}$, from which the presence or absence of quantum chaos is inferred. The protocol consists of measuring observables in a single-qubit randomized basis, then classically learning the entanglement Hamiltonian $H_\text{E}(\{\beta_i\})$, which parameterizes the reduced quantum state with parameters $\{\beta_i\}$, so as to  optimally reproduce all measurements. The statistical behavior of the eigenvalue spectrum of the entanglement Hamiltonian is then analyzed: e.g., eigenvalue repulsion indicates quantum chaos, as detailed in the main text.
    } 
	\label{fig:introoverview}
\end{figure*}

Gauge theories and their lattice formulations are among prime physical models whose simulations will benefit from quantum-computing technology~\cite{Banuls:2019bmf,Klco:2021lap,bauer2023quantum,Bauer:2023qgm,di2023quantum}. Gauge theories are key in high-energy and nuclear physics~\cite{aitchison2012gauge,quigg2021gauge}, condensed and synthetic quantum matter~\cite{fradkin2013field,kleinert1989gauge,wen1990topological,levin2005string}, local fermion-to-qubit mappings~\cite{chen2018exact,chen2020exact,chen2023equivalence}, and quantum-error correction ~\cite{kitaev2003fault,kitaev2006anyons,sarma2006topological,nayak2008non,lahtinen2017short}. Studying thermalization dynamics of gauge theories, e.g., in early universe and in high-energy particle collisions, remains challenging using first-principles simulation methods~\cite{berges2021qcd}. As a first step in experimentally probing thermalization dynamics of gauge theories, we study a $Z_2$ lattice gauge theory (LGT) in $2+1$ spacetime dimensions~\cite{maunz2016high,monroe2021programmable} using a digital trapped-ion quantum computer~\cite{pogorelov2021compact,monroe2021programmable,yao2022experimental,bohnet2016quantum}. 
We use a chain of fifteen $^{171}$Yb$^+$ ions to realize a general-purpose fully-connected digital quantum computer with twelve qubits, schematically shown in Fig.~\ref{fig:introoverview}(a), and use this computer to natively and accurately encode system's initial state, evolve this  state in time, and measure observables.

Our analysis relies on EH tomography~\cite{pichler2016measurement,dalmonte2018quantum,kokail2021entanglement,kokail2021quantum,zache2022entanglement,mueller2023quantum} in combination with randomized-measurement protocols~\cite{ paini2019approximate,huang2020predicting, huang2021efficient,hu2021classical, kunjummen2021shadow, levy2021classical, huang2021demonstrating,huang2022learning,
bringewatt2023randomized,elben2019statistical, elben2020mixed,elben2023randomized} to learn representations approximating non-equilibrium states. The EH is defined as $H_{\rm E} \equiv -\log(\rho_A)$, where $\rho_A$ denotes the reduced density matrix of subsystem $A$ formed by bipartitioning a (pure) quantum state. The utility of EH as a theoretical and experimental tool stems from the observation that, in many cases, it consists of approximately local operators~\cite{bisognano1975duality,bisognano1976duality}. This resemblance to conventional Hamiltonians, i.e., energy operators governing system dynamics, facilitates simple theoretical analysis: If an EH is ($k$-)local, it can be described by parameters whose number scales polynomially with the system size, unlike the matrix $\rho_A$ which requires an exponential number of parameters.

Randomized-measurement protocols~\cite{elben2023randomized} can be used to learn the EH: after repeatedly preparing and then measuring the quantum state in a randomly chosen basis, one can fit a  parameterized EH to the obtained measurement outcomes.
Because the spectrum of $H_{\rm E}$, known as the entanglement spectrum, scales logarithmically with the Schmidt eigenvalues of $\rho_A$, it is difficult to learn an EH and reproduce the {entire} entanglement spectrum with arbitrary accuracy. Randomized-measurement protocols may require resolving exponentially small probabilities even for relatively small systems, which is out of reach of current quantum devices and perhaps fault-tolerant devices, too. 
The present experiments can, therefore, only aim at approximating $H_{\rm E}$ and accurately reproducing its low-energy spectrum. Applications of this approach include the verification of topological phases~\cite{mueller2022thermalization,bringewatt2023randomized}.

Nonetheless, what if the precise quantitative structure of $H_{\rm E}$ were not crucial, but only its {statistical properties} were? This viewpoint is that of random matrix theory~\cite{guhr1998random,mehta2004random}, where metrics like level distribution and spectral form factor differentiate between integrable and chaotic dynamics, while being indifferent to quantitative details. For instance, a model's symmetries influence spectral correlations of the EH and must, therefore, be taken into account~\cite{rosenzweig1960repulsion,giraud2022probing}. This statistical perspective underpins our study. Guided by physical insights regarding the expected operator content of a non-equilibrium EH, we ask whether one can learn, from experimental data, a classical representation of a state to answer a simpler question: does a quantum state exhibit {universal} signatures of quantum chaos evident in the statistical properties of its EH? Crucially, can we discern this scenario from one where the state lacks chaotic behavior?

To answer these questions, we focus on two observables indicative of quantum ergodic and chaotic behavior: the entanglement(-Hamiltonian) gap-ratio distribution (EGRD)~\cite{oganesyan2007localization,chang2019evolution,rakovszky2019signatures} and the entanglement spectral form factor (ESFF)~\cite{daug2023many,ma2022quantum}. The EGRD is predicted to exhibit level repulsion for chaotic states, in contrast to uncorrelated levels in non-chaotic scenarios. Similarly, the ESFF is predicted to display a plateau-ramp structure in chaotic states.
~{Strictly speaking, both quantities indicate quantum ergodicity which usually implies quantum chaos.} To experimentally test these predictions, we initialize our system in a product state and perform a quantum quench by digitally evolving the system for a variable length of time. Our protocol to access these quantities is depicted schematically in Fig.~\ref{fig:introoverview}(b). At short evolution times, we observe an EGRD that is characteristic of uncorrelated states. With increasing evolution time, we observe the onset of level repulsion and a ramp-plateau structure in the ESFF, indicative of quantum chaos. Detailed analysis, including comparison with emulated data, suggests that the observed behavior primarily arises from the thermalization dynamics of the isolated quantum system under simulation, with experimental inaccuracies and shortcomings of our ansatz playing a small, though non-negligible, role. Our work, therefore, establishes present-time quantum computers as robust tools for studying universal feature of thermalization dynamics in complex many-body systems.

\section*{Results}
\noindent
We focus on a $Z_2$ LGT in $2+1$ D with the Hamiltonian
\begin{align}\label{eq:Z2Hamiltonian}
    H = \sum_\square W_\square +g \sum_\ell \sigma_\ell^z\,.
\end{align}
Here, $W_\square\equiv \prod_{\ell \in \square} \sigma_\ell^x$ is a magnetic-field operator, where $\square$ denotes the elementary (square) plaquettes of the lattice, see the Methods section. $\sigma^x_\ell$ and $\sigma^z_\ell$ are Pauli operators representing gauge link and electric fields on edge $\ell$, respectively. These act on spin-$\frac{1}{2}$ hardcore bosons residing on the edges of a two-dimensional spatial square lattice. The first term in \Eq{eq:Z2Hamiltonian} corresponds to the magnetic energy, and the second to the electric energy, with coupling $g$ controlling the relative strength of the two non-commuting contributions. In this formulation, the $z$ basis, therefore, represents the electric basis, while the $x$ basis corresponds to the magnetic basis. 
The Hamiltonian is expressed in dimensionless units and $\hbar$ is set to one throughout.

\begin{figure}[t]
  \centering
	\includegraphics[width=0.485\textwidth]{
 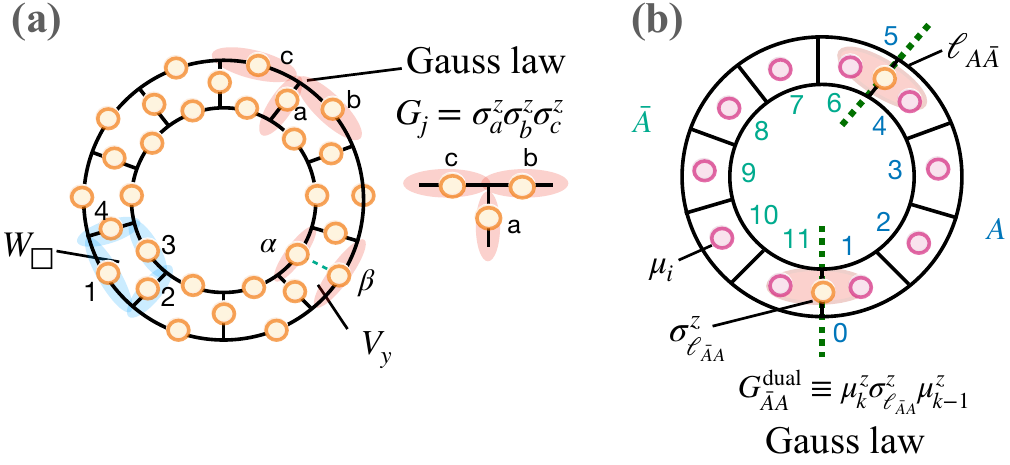}
	\caption{
\textit{Model overview.} 
 (a) The spatial lattice considered in this work consisting of $L_x = 10$ plaquettes along the $x$ direction with periodic boundary conditions and $L_y=1$ plaquette along the $y$ direction with fixed boundary conditions. Degrees of freedom (DOFs) are  (spin-$\frac{1}{2}$ hardcore bosons denoted by orange circles) residing on the edges of a two-dimensional square array. The electric-field (link) operators on each edge are denoted by pink (blue) ellipses. Electric operators at site $j$ build the Gauss-law operator $G_j$, and the four link operators covering edges of a plaquette build the magnetic operator $W_\square$. The short-ribbon operator $V_y$, comprised of electric-field operators at the adjacent links $\alpha$ and $\beta$, is also shown. b) The lattice in the dual formulation. Gauge-dependent operators (e.g., $\sigma^z_{\ell_{\bar{A}A}} $) and the corresponding Gauss laws (e.g., $G_{\bar{A}A}^{\rm{dual}}$ denoted by the pink ellipsoid) on the boundary between the system  $A$ and its complement $\bar{A}$ (indicated by green thick dashed lines) remain unchanged from the original formulation. However, bulk operators within each subsystem are replaced with gauge-independent Pauli operators $\mu_{i}^{x,z}$ on each plaquette. The 12 quantum DOFs in this system are mapped to the trapped ion quantum computer, with DOFs shown as pink circles mapped to qubits 1 to 4 in the system and qubits 6 to 11 in the complement. Boundary DOFs, shown as yellow circles, are mapped to qubits 0 and 5.
}
	\label{fig:model}
\end{figure}

We consider the joint +1 eigenspace of the Gauss-law operators $G_{j}\equiv \prod_{\ell \in +_j} \sigma^z_\ell$ as the physical Hilbert space, where $\ell \in +_j$ denote the 
links adjacent to a lattice site $j$. In the subsequent discussion, we consider a quasi-1D chain composed of $L_x$ plaquettes along the $x$ direction with periodic boundary conditions. Fixed boundary conditions are applied along the short side of the chain (with $L_y=1$ plaquettes) as illustrated in \Fig{fig:model}
(a). In this configuration, all Gauss laws contain three links at each site. Consider a \emph{short-ribbon} operator $V_{y}\equiv \sigma^z_\alpha\sigma^z_\beta$, where $\alpha$ and $\beta$ are the two opposing links along the inner and the outer circumference {shown in \Fig{fig:model}}. {The position of $\alpha$ and $\beta$ is arbitrary, and $V_y$ is unique: different choices of $\alpha$ and $\beta$ are related to each other upon application of the Gauss-law operator under which states are invariant} 
 This operator commutes with the Hamiltonian and defines superselection sectors of the model; we work in the $V_y=1$ sector (i.e., short-ribbon operators possess eigenvalue one).

\begin{figure*}[t]
  \centering
	\includegraphics[width=0.95\textwidth]{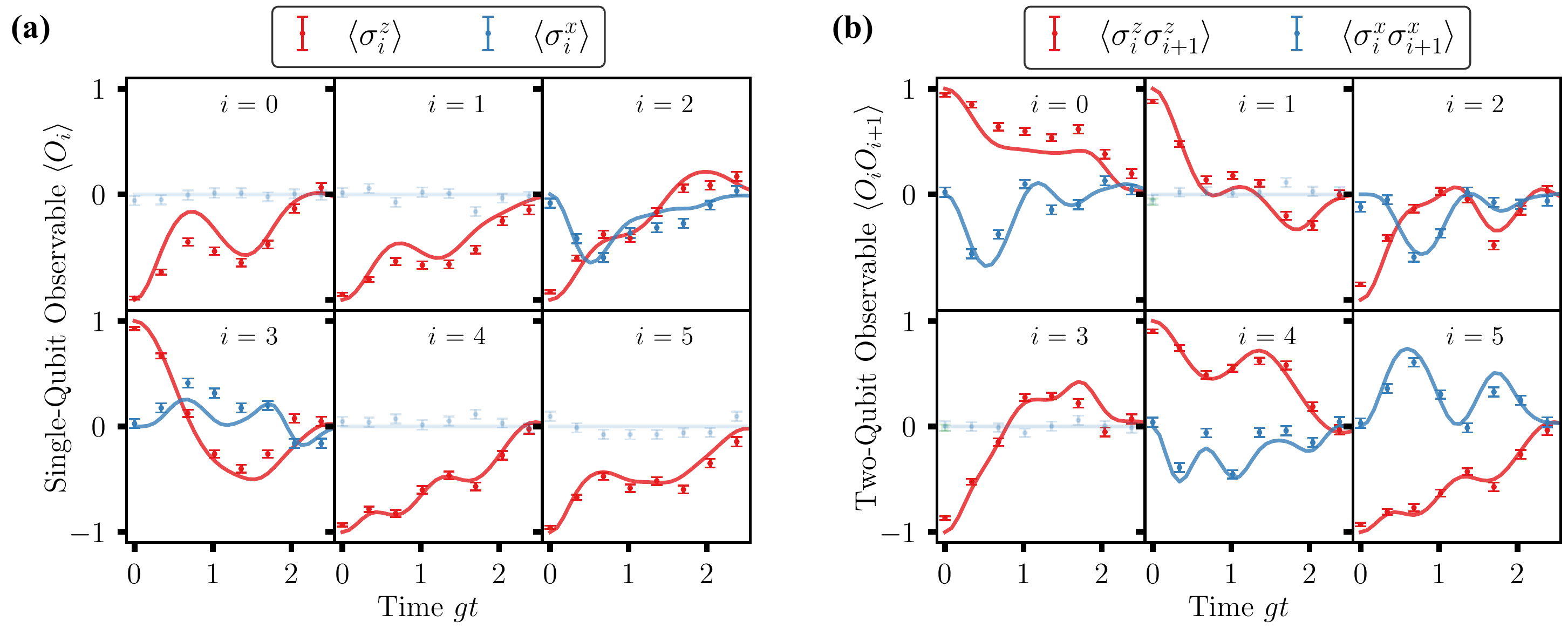}
	\caption{\textit{Real-time observables in $Z_2$ LGT in (2+1)D.} Measured expectation values of single-qubit (a) and two-qubit (b) observables of qubits $i=0,\dots,5$ of the 10-plaquette theory with $g=0.85$ following four steps of Trotterized time evolution with an initial state $\ket{\Psi_0}=\ket{\downarrow\downarrow\downarrow\uparrow\downarrow\downarrow\uparrow\uparrow\uparrow\downarrow\uparrow\uparrow}$. Each point represents data from $N_{\rm shots} = 500$ experiments. {Error bars indicate statistical uncertainties obtained using $N_{\rm boot}$ = 1000 bootstrap samples and do not include systematic errors.} Red (blue) points correspond to measurements in the Bloch $z$-basis ($x$-basis); light-blue points indicate the expectation values of non-gauge--invariant observables, which are expected to be zero. Solid lines correspond to exact classically-computed results. }
	\label{fig:experimentbasis}
\end{figure*}

For the thermalization study, we will bipartition the system to subsystem $A$ and its complement $\bar{A}$, resulting in two boundaries, which we call $\ell_{A\bar{A}}$ and $\ell_{\bar{A}A}$. To directly encode the degrees of freedom of the model described onto the qubits, one requires one qubit per link degree of freedom, or $N_q=3L_x$ qubits in total. 
To adapt the simulation to the available hardware, we employ a dual formulation of the model, described in the Methods section, that requires only $ N_q=L_x + 2 $ qubits. The dual formulation maps all bulk operators onto gauge-invariant ones ($\mu_i$), except at the two boundaries where the original operators are retained, as shown in Fig. \ref{fig:model}
(b). This mapping guarantees that the entanglement structure---the primary observable in our study---remains identical between the dual and the original lattice gauge theory. 
Similarly, both the original and the dual theory exhibit four symmetry sectors, which are introduced in the Methods section.

\subsection*{Digital quantum computation of real-time evolution}
We first demonstrate the capability of our digital computer to compute the dynamics of the $Z_2$ LGT model, including initial-state preparation, time evolution, and subsequent measurements.

We initialize the system in a randomly chosen electric eigenstates that respects Gauss law ($\ket{\Psi_0}=\ket{\downarrow\downarrow\downarrow\uparrow\downarrow\downarrow\uparrow\uparrow\uparrow\downarrow\uparrow\uparrow}$, see the Methods section,
and simulate its time evolution. Explicitly, we approximately obtain the state $\ket{\psi(t)}=U(t)\ket{\Psi_0}$ with $U(t)=e^{-itH^{\rm{dual}}}$, via a Trotterization protocol. Each Trotter step separately implements the non-commuting terms of $H^{\rm{dual}}$, i.e., those diagonal in electric ($z$) or magnetic ($x$) bases. We implement four Trotter steps and set the coupling to $g=0.85$. Our choice of coupling is such that magnetic and electric terms are of similar magnitude, making the model sufficiently non-integrable.  Since the initial state is not an eigenstates of the full Hamiltonian, the system undergoes non-trivial time evolution when evolved under $H^\text{dual}$. For each  evolution time $t$, we apply single-qubit gates to measure all qubits in either the Bloch $x$ or the $z$ bases and repeat each experiment $N_{\rm{shots}}=500$ times. 
Quantum circuits implementing initial-state preparation, Trotterized time evolution, and measurements are provided in the Methods section. Each Trotter step of evolution uses 12 variable-angle Mølmer-Søerensen (MS) gates.

\begin{figure*}[t]
  \centering
	\includegraphics[width=0.95\textwidth]{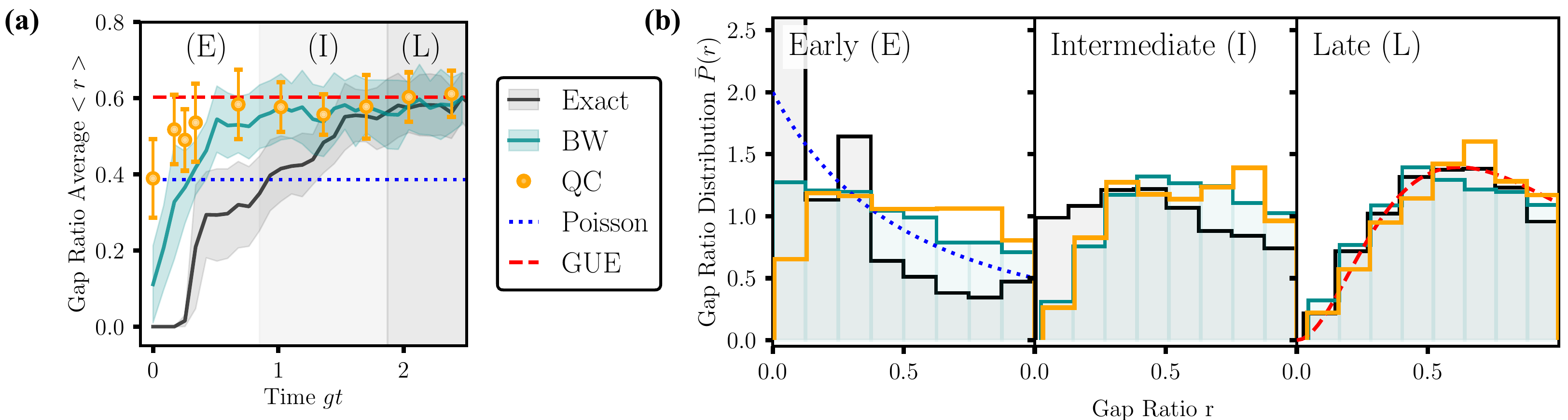}
	\caption{ \textit{Statistics of the gap ratios of the spectrum of the entanglement Hamiltonian.} (a) Time evolution of the average gap ratio averaged over 6 randomly-drawn initial states and all symmetry sectors. The horizontal lines represent the averages for non-repulsive (Poisson, blue dotted) and repulsive (GUE, red dashed) distributions. Error bars indicate standard deviation of the mean over the initial states and the symmetry superselection sectors of the reduced density matrix (see the Methods section). (b) Distribution of the entanglement-spectrum gap ratios, combined across 6 randomly-drawn initial states, all symmetry sectors, and all times in each of the regimes {Early (E), Intermediate (I), and Late (L)}. A total of 840, 504 and 336 gaps are quantum-computed {based on $10^6$ total shots}, and their average is shown in orange. Simulated Bisognano-Wichmann results in the limit of infinite measurements are shown in cyan, and the exact predicted distributions in black. Blue-dotted and red-dashed curves represent Poisson and GUE distributions.
    }
	\label{fig:gap-ratio-exp-analysis}
\end{figure*}

The measurement results are used to compute single- and two-qubit observables in the Bloch $x$ and $z$ bases and are plotted in Figure~\ref{fig:experimentbasis} for select observables. The statistical errors are determined by a standard bootstrap resampling analysis with $N_\text{boots}=1000$ bootstrap samples. {We have verified that the our results are 
    independent of $N_{\rm boot}$.}
    {The expectation values of the physical operators (dark blue and red) follow the zero-parameter theory prediction (solid lines). The observed deviations are well-explained by $\sim$10\% systematic under-rotations of the XX gates, likely caused by calibration drift (see Supplementary Figure S2).} Importantly, the expectation values of gauge-invariance--violating operators (i.e., those that do not commute with the Gauss laws at the two boundaries), shown in light blue, are consistent with zero within measurement errors. In Section 1 of the Supplementary Information, which includes Refs.~\cite{mueller2022thermalization,ghosh2015entanglement,van2016entanglement,huang2021demonstrating}, gauge-invariance violation is shown to be about 5\% throughout the time evolution. { The approach of observables to zero at late times is a coincidence associated with the time frame chosen---it neither signals thermalization nor decoherence. Additionally, we have verified robustness at larger gate angles by collecting data at even later times but using the same number of Trotter steps, see the Supplemental Figure 3. 
    }

\subsection*{Thermalization Dynamics of $Z_2$ Lattice Gauge Theory with Trapped Ions}
To study thermalization, we randomly initialize the system in Gauss-law--respecting electric eigenstates. 
We then define a subsystem $A$ consisting of $L_A \le L_x/2$ plaquettes. The reduced density matrix $\rho_A(t)$ of $A$ after evolution time $t$ is related to the density matrix $\rho(t)$ of the whole system as
\begin{align}\label{eq:defreducedDM}
    \rho_A (t) \equiv \text{Tr}_{\bar{A}}[ \rho(t)], \quad  \rho(t)\equiv U(t) \rho_0 U(t)^\dagger,
\end{align}
where $\rho_0 \equiv |\Psi_0 \rangle\langle \Psi_0 |$.
Our main interest lies in the statistical properties of the Schmidt decomposition of the reduced density matrix of the subsystem $A$, given by $\rho_A(t) \equiv \sum_\lambda p_\lambda(t) | \lambda(t) \rangle \langle \lambda(t)| $, and its associated EH~\cite{li2008entanglement,dalmonte2022entanglement}
\begin{align}\label{eq:entHamiltonian}
H_{\rm E}(t) \equiv -\log(\rho_A(t))\,.
\end{align}
The EH's eigenspectrum, called the entanglement spectrum, is defined by $\{\xi_\lambda(t) \equiv -\log(p_\lambda(t))\}$.

The statistical properties of the entanglement spectrum directly inform the thermalization dynamics of the system. One statistical probe is the distribution, $P(r)$, of the gap ratios, $r_\lambda$, of the entanglement spectrum~\cite{oganesyan2007localization},
\begin{align}\label{eq:gapratiodef}
    r_\lambda \equiv \frac{ \min(\delta_\lambda , \delta_{\lambda-1})}{\max(\delta_\lambda , \delta_{\lambda-1})}\, ,
\end{align}
where $\delta_\lambda \equiv \xi_\lambda - \xi_{\lambda-1} \ge 0$ are the gaps between the eigenvalues $\xi_\lambda$ of $H_{\rm E}$ and $r \equiv \{r_\lambda \}$. According to random matrix theory of Hermitian matrices, this quantity distinguishes chaotic from nonchaotic behavior depending on whether the distribution is centered away from zero (level repulsion) or centered around zero (uncorrelated levels)~\cite{guhr1998random}. 

In classical computation of a tractably small system, the state and its entanglement are readily available. In a quantum-simulation experiment, however, these must be inferred from measurements. To tackle this challenge, we constrain the operator content of the EH to obtain an approximate $H_\text{E}$, and utilize an EH-tomography scheme based on randomized measurements~\cite{pichler2016measurement,dalmonte2018quantum,kokail2021entanglement,kokail2021quantum,zache2022entanglement,mueller2023quantum}.
Concretely, we perform a single-layer, single-qubit randomized-measurement~\cite{brydges2019probing} of the time-evolved state. 
At the end of the time evolution, this protocol applies one of $N_\mathcal{U}$ different gates $\mathcal{U}_j = \otimes_{i=0,\cdots,N_q-1} u_{i,j}$ for $j=1,\cdots,N_\mathcal{U}$ consisting of independent single-qubit gates $u_{i,j}$ sampled from a unitary 2-design~\cite{dankert2009exact}. For each $\mathcal{U}_j$, $N_{\rm{shots}}$ experiments are performed, measuring all qubits in the Bloch $z$ basis. The relative frequencies of the different bitstrings are then compared with the prediction based on an approximate EH inspired by the Bisognano-Wichmann theorem~\cite{bisognano1975duality,bisognano1976duality}. Details are provided in the Methods section.

To study the evolution of the entanglement spectrum in the $Z_2$ LGT, we repeat the time-evolution experiments shown in Fig.~\ref{fig:experimentbasis} starting from 6 randomly chosen initial electric eigenstates that satisfy the Gauss laws. 
We numerically checked that the results remain valid for other randomly chosen states in this basis. 
For each evolution time $t$, symmetry sector, and initial state, we subsequently perform randomized-measurement tomography with $g=0.85$, $N_\mathcal{U}=24$ bases, and $N_{\rm shots}=750$ bitstring measurements in each basis, and use these measurements to reconstruct the EH as described in the previous section. This procedure yields a set of gap ratios for each initial state, symmetry sector, and evolution time.

In \Fig{fig:gap-ratio-exp-analysis}(a), we plot in orange points the average gap ratio $\langle r\rangle \equiv \sum_r r\bar{P}(r)$ of the reconstructed EH  as a function of the scaled evolution time $g t$. The plotted gap ratios are averaged over both the symmetry sectors and the 6 randomly chosen initial states. The black lines correspond to the predicted exact distributions following the Trotterized time evolution. The cyan lines correspond to the EH obtained from an optimal BW-inspired ansatz, where we numerically minimize the
the relative entropy (Kullback–Leibler divergence) between the exact state and the ansatz, corresponding to the limit of infinitely many measurements. A buildup of the level repulsion is discernible as the observed average gap ratio $\langle r \rangle$ evolves from $\approx 0.4$ predicted for a non-repulsive Poisson distribution (blue dashed line) towards $\approx 0.6$ characteristic of repulsive level statistics of a Gaussian Unitary Ensemble (GUE)~\cite{oganesyan2007localization}. Three time regimes {Early (E), Intermediate (I), and Late (L)} are identified that correspond, respectively, to the evolution of the predicted average gap ratio toward the Poisson-distribution value, toward the GUE value, and to saturation at the GUE value. {These regimes were determined 
before taking data, based on the numerically computed results (black solid lines in \Fig{fig:gap-ratio-exp-analysis}), hence constituting a prediction.} 

In \Fig{fig:gap-ratio-exp-analysis}(b) we plot, for the three ranges of evolution times, the corresponding normalized distribution of the gap ratios, $\bar{P}(r)$, combined over 6 random initial states and 4 symmetry sectors. We observe a clear transition from early-time non-repulsion (Poisson distribution, blue dotted line) {in the early regime (E) to level repulsion (GUE, red) in the late regime (L).} At intermediate times, a distribution is observed between the initial absence of level repulsion and the subsequent emergence of level repulsion.  Error bars and bands denote the variance resulting from averaging over symmetry sectors and initial states. The predicted exact distribution (black line) exhibits a sharper peak around zero at the earliest times due to the reduced density matrix not being full rank. Conversely, the BW-inspired ansatz (cyan line) generally parametrizes a full-rank matrix unless couplings are very finely tuned, resulting in an overestimation of level repulsion in the initial stages. 
{The time scale required to achieve a distribution consistent with GUE predictions is approximately the same for all randomly-chosen initial states. This time scale is expected to scale with subsystem size as was first observed in Ref.~\cite{rakovszky2019signatures} in a (1+1)D spin model, see also Supplemental Figure 5.}

 {The main effects of Trotterization versus exact time evolution are two-fold. First, the evolution is governed by  an effective instead of the exact Hamiltonian; however, we draw our BW parameterization ansatz from the exact Hamiltonian. Second, at later times and with large Trotter-step sizes (not shown here), a recurrence of the initial state is observed, and the gap-ratio statistic becomes `un-chaotic' again, see Supplemental Figure 8.}
\begin{figure}[t]
  \centering
	\includegraphics[width=0.49\textwidth]{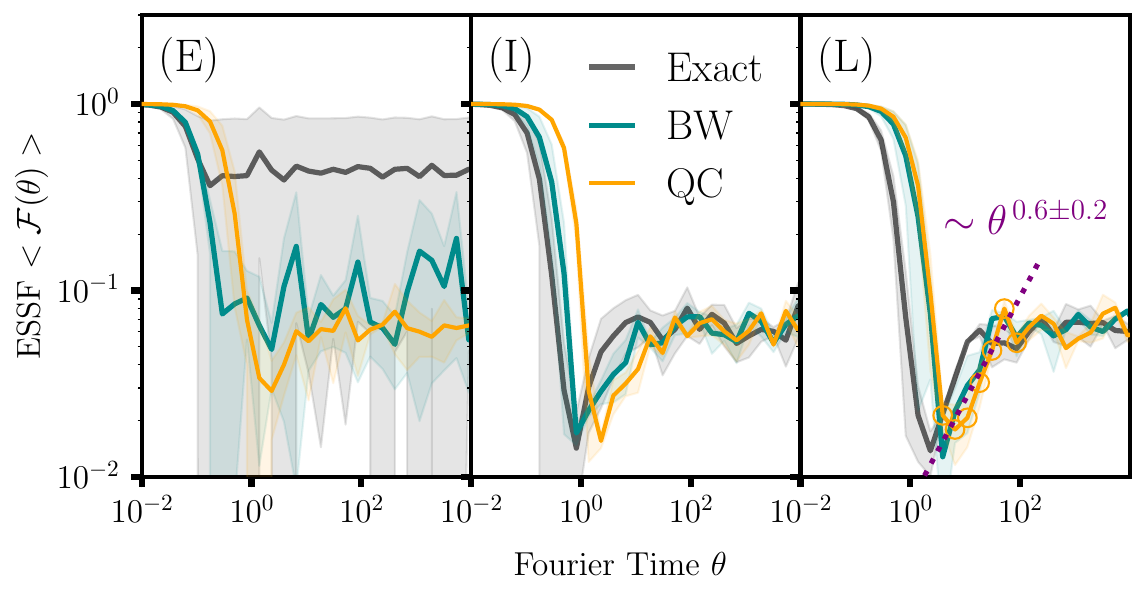}
	\caption{\emph{Entanglement spectral form factor}. The average entanglement spectral form factor across various initial states, symmetry sectors, and three distinct regimes {Early (E), Intermediate (I), and Late (L)} identified in Fig.~\ref{fig:gap-ratio-exp-analysis}. The figures display exact results (black curves), infinite-measurement outcomes (cyan), and quantum-computed experimental data (orange). In panel (L), a purple dotted line indicates a power-law fit of the ramp observed in the quantum-computed data. Our normalization ensures $\langle \mathcal{F}(0)\rangle =1$, with the plateau occurring at $\langle \mathcal{F}(\infty)\rangle =1/ d_s$, where $d_s$ denotes the dimension of a symmetry block, see the Methods section for details. Shaded areas indicate the standard deviation over initial states, symmetry sectors, and times. {Data included in the fit are shown as circles.} }
	\label{fig:ESFF-simulator-analysis}
\end{figure}

A complementary view of the statistics of the entanglement spectrum that captures global correlations in the level distribution is afforded by the entanglement spectral form factor (ESFF)~\cite{guhr1998random,chang2019evolution},
\begin{align}\label{eq:defESFF}
    \mathcal{F}(\theta ;t) \equiv \Big\langle \frac{1}{\mathcal{R}^2_A(t)} \sum_{\lambda,\lambda'} e^{i\theta[\xi_\lambda(t) -\xi_{\lambda'}(t) ]} \Big\rangle\,,
\end{align}
where $\langle \cdot \rangle$ denotes the average over initial states $\ket{\Psi_0}$. Here, $\mathcal{R}_A \equiv \lim_{\alpha \rightarrow 0} \exp\{ \frac{1}{1-\alpha} \log(\sum_\lambda p_\lambda^\alpha)\}$~\cite{chang2019evolution}  is the effective rank of $H_{\rm E}$, whose value depends on the state and lies in the range $[1, 2^{L_A}]$. Only the $\mathcal{R}_A$ lowest levels of the EH are included in the analysis. {Small values of $\theta$ probe global correlations among eigenvalues which  are not universal, while large values  probe correlations between eigenvalues that are progressively closer and where GUE proedictions apply. } 
In \Fig{fig:ESFF-simulator-analysis}, we plot the ESSF reconstructed from the data used in \Fig{fig:gap-ratio-exp-analysis}. The three panels of the plot correspond to the three time regimes discussed in relation to the EGRD evolution. The displayed  curves are averaged over initial states, symmetry sectors, and each of the three time ranges. Starting from an initial flat behavior of the ESSR as a function of $\theta$ in panel (E), our data in panels (I) and (L) clearly show the buildup of a ramp-plateau structure, indicating ergodic behavior and implying quantum chaos~\cite{chang2019evolution}. In panel (L), we show a fit (purple dotted line) to the observed ramp, indicating a power-law behavior $\theta^\kappa$ where $\kappa = 0.6 \pm 0.2$, with the fit error determined by changing the fit regime. {The exponent depends on the definition of the ESSF in the presence of symmetries. Our definition leads to a value that is smaller than the one from RMT.}  It is consistent with the results obtained from the numerical analysis of a significantly larger  system in Supplementary Information sec.~2.

Our results demonstrate that, using observables that reveal universal statistical or global features of the entanglement spectrum, the onset of chaotic behavior can be robustly traced in the entanglement dynamics of arbitrary initial states. Importantly, time digitization and experimental infidelities in current systems do not significantly impact qualitative features of thermalization dynamics.

\section*{Discussion
\label{sec:conclusions}}
\noindent
Using a digital trapped-ion quantum computer, we observe early-stage thermalization dynamics in a $Z_2$ lattice gauge theory in 2+1 dimensions by generating nonequilibrium quantum states and measuring their entanglement structure. As performing quantitative state tomography is extremely challenging and resource intensive, we focused on a simpler question: \emph{can a randomized-measurement--based entanglement-Hamiltonian
tomography recover universal properties of the EH that indicate quantum chaos?} 

{The term `universal' refers to the fact  that, on timescales of order one in units of inverse coupling, the entanglement spectrum of thermalizing systems develops correlations between neighboring levels that are described by a universal (GUE) statistical ensemble, independent of the initial conditions.}
{For integrable systems, the gap ratio may not exhibit universal behavior. While a Poisson distribution emerges in certain cases~\cite{froland2024entanglement}, the distribution generally remains sub-Poissonian, may oscillate, or can be ill-defined if the rank of the reduced density matrix remains small throughout the evolution.}

We experimentally determine the entanglement spectrum of time-dependent quantum states and use two properties of this spectrum, namely the entanglement gap-ratio distribution [Eq.~\eqref{eq:gapratiodef}] and the  entanglement spectral form factor [Eq.~\eqref{eq:defESFF}], as indicators of the emergence of quantum chaos. Our data indicate that both quantities behave as expected: the initially non-repulsive level distribution transitions to a repulsive one. Likewise, a ramp-plateau feature of the ESFF is initially absent but builds up with time. {The emergence of quantum chaos in the EH is a prerequisite for thermalization which takes parametrically longer time to set in. This separation is discussed in Ref.~\cite{rakovszky2019signatures}: while the timescale for the EH to become chaotic is governed by the Lieb-Robinson velocity describing local-operator spread, entanglement saturation---and thus thermalization---is much slower~\cite{mezei2017entanglement,jonay2018coarse}.}

The key technique enabling our analysis is a Bisognano-Wichmann--inspired ansatz for the EH, which allows the parameterization of a nonequilibrium state using a polynomial number of parameters. While this classical parameterization optimally reproduces the observed (randomized) measurement results, in Supplementary Information~2, we show that the higher-lying part of the EH spectrum is not quantitatively recovered. Indeed, the measurement cost for {precise} state tomography still scales exponentially with the subsystem size. Fortunately, to detect the presence of quantum chaos, one needs to only distinguish repulsion from non-repulsion in the EGRD  or identify a ramp-plateau structure in the ESFF. Our work indicates that selecting a sufficiently small subsystem of a much larger, potentially classically nonsimulable system, can sufficiently constrain these observables. 

{Other signatures of thermalization can also be examined, such as the saturation of local observables or the entanglement entropy to their thermal values. However, beyond the longer simulation times required, several caveats must be considered when using such quantities to assess thermalization. For instance, finite-volume effects affect local observables at late times, and, and certain local observables do not accurately distinguish thermalization from integrable dynamics; see, e.g. Ref.~\cite{rigol2007relaxation}.}

Our randomized-measurement procedure remains classically simulable because the lattice sizes we considered are manageable with exact diagonalization. Our procedure is also heavily tailored at minimizing the computational load on the quantum computer, avoiding deep quantum circuits and the need for error mitigation, at the cost of classical post-processing. To extend our approach to larger systems, several steps can be implemented in future work:
\begin{itemize}
{
\item[$\circ$]
Our study focuses on gauge theories. As demonstrated in the example studied, gauge theories possess a highly intricate Hilbert-space structure, shaped by an extensive set of local constraints, i.e., Gauss's laws, that govern their dynamics. These constraints also give rise to a non-trivial entanglement structure. For example, resolving the symmetry properties of the reduced density matrix was crucial to identifying the presence of quantum chaos. Without knowledge of this structure, simply computing the gap-ratio distribution of all eigenvalues of the EH would have shown a Poisson, or more singular distribution, falsely implying the absence of quantum chaos because eigenvalues from different symmetry blocks are  uncorrelated. It would be interesting to extend this study to continuous groups, as well as non-Abelian gauge theories, which can pave the way toward studying thermalization physics of the Standard Model.}

\item[$\circ$]
The employed single-qubit one-layer randomization strategy is symmetry ignorant, i.e., it randomizes regardless of the known symmetry structure of the subsystem density matrix $\rho_A(t)$. Although our scheme is tomographically complete, and enables state reconstruction along with its symmetries, it is inefficient.  Symmetry-conscious protocols, for the LGT in this work and for other models, have been developed~\cite{bringewatt2023randomized,hearth2023unitary,hearth2023efficient}, { providing tomographically complete circuits with polynomial depth}, and could be adopted in future work.

\item[$\circ$]
A major methodological uncertainty is the BW-inspired parameterization of the EH, which is entirely heuristic. While there is research exploring the operator content of the EH of ground, excited, and thermal states~\cite{joshi2023exploring}, further investigations are needed into the applicability and limitations of EH-based schemes for far-from-equilibrium states. Generally, the BW(-inspired) ans\"{a}tze describe the \emph{low-energy} regime of the EH well, and incorporating progressively more nonlocal terms  improves convergence into the bulk~\cite{kokail2021entanglement} which the EGRD, and to a lesser extent the ESFF, predominantly rely on.  Notably, our analysis reveals that despite the quantitative discrepancy in the bulk, the statistical distribution of the EH appears to be accurately reproduced. This suggests that it may be unnecessary to perform {precise} state tomography when one's interest is solely in statistical properties. 

\item[$\circ$]
Trotterization corresponds to time evolution with an effective, rather than the desired, Hamiltonian. We observe that employing too few Trotter steps leads to poor convergence of our optimization procedure for the BW-inspired ansatz. In this case, more nonlocal operators must be included in the ansatz; constraining these becomes challenging, especially when relying on a limited number of measurements. Since using a large number of Trotter steps increases 
experimental errors, the optimal Trotterization and its interplay with the BW ansatz should be further investigated.

\item[$\circ$]
Detailed comparison between the emulator and the experimental data (see Supplementary Information) highlights the influence of device errors, primarily for the time evolution and tomography steps of our algorithm. {Developing a comprehensive error model to predict how errors impact the BW-EHT analysis for fully digital computations is a complex task that we have not investigated in detail. However, this question has been explored  in Ref.~\cite{kokail2021entanglement} for an analog-digital scheme; a comprehensive error model for the specific computing platform used in our work can be found in Ref.~\cite{debroy2020logical}.} The dominant errors in our study are $Z$-flip errors and under-rotations due to mechanical motion of the ions~\cite{cetina2022cot}.
Our tomography protocol is especially susceptible to small single-qubit rotation errors during randomization~\cite{brydges2019probing}.  
Recently demonstrated sympathetic cooling during circuit execution \cite{cetina2022cot} would allow the needed measurement circuits to be executed with higher fidelity. The same technique could extend our study to larger systems, later times, and more Trotter steps, e.g., allowing us to directly test the eigenstate thermalization hypothesis in a digital quantum-computing set-up. 
{Once system sizes exceed classical simulability, verification becomes an important problem; for an overview see, for instance, Ref.~\cite{carrasco2021theoretical}.}
\end{itemize}

In summary, our results demonstrate that entanglement structure is a measurable quantity in present-day LGT quantum-simulation experiments, and illustrates the potential value of our approach to probe thermalization dynamics and its robust universal features in strongly coupled isolated quantum many-body systems. A compelling future direction is to extend the investigation to later times, to probe aspects such as pre-thermalization~\cite{berges2004prethermalization,gring2012relaxation,mori2018thermalization, kyprianidis2021oop,neyenhuis2017oop} or fluctuation-dissipation relations~\cite{callen1951irreversibility,kubo1966fluctuation,orioli2019breaking,schuckert2020probing}, once experimental capabilities permit. 
This would also allow for probing the applicability of the Eigenstate Thermalization Hypothesis, e.g., in systems with non-Abelian symmetries~\cite{murthy2023non} and other gauge theories~\cite{Yao:2023pht,ebner2024eigenstate}. Quantum many-body scars have been identified in a slightly different lattice geometry of this model and may serve as a means of verifying quantum simulation experiments beyond the reach of classical computation~\cite{hartse2024stabilizer}. Furthermore, the experimental and theoretical tools of this study can be applied in a number of other applications, including obtaining thermodynamic quantities such as work and heat exchanged during nonequilibrium processes~\cite{davoudi2024quantum}, and detecting phases of matter, including topological phases~\cite{bringewatt2023randomized}, in quantum-simulation experiments. {Ultimately, strategies such as those presented in this work can reveal the role of entanglement in QCD thermalization, and may shed light on the short thermalization time scales observed in heavy-ion collisions~\cite{berges2021qcd}. Such studies, nonetheless, will require large-scale fault-tolerant quantum computers. In the meantime, studies in simpler gauge theories, such as in confining lower-dimensional models, can potentially provide the first clues.}

\section*{Methods}
In this section, we provide further details on the model studied in this work, our experimental trapped-ion setup, the entanglement-Hamiltonian tomography we employed, and the quantum-circuit representation of our full quantum-simulation protocol.

\subsection*{Generalized Ising Duality}
To adapt the simulation to the available hardware, we consider a dual formulation of the model that requires only $ N_q=L_x + 2 $ qubits. In the dual formulation, all operators, except those containing operators having support  at the boundaries between the subsystems, labeled $\ell_{\bar{A}A}$ and $\ell_{A\bar{A}}$, are entirely represented using gauge-invariant (Ising dual) spin-$\frac{1}{2}$ variables which act solely within the physical Gauss-law subspace; they are denoted by $\mu_i^{x,z}$ Pauli matrices [see purple circles in \Fig{fig:model}
(b)]. Concretely, $\mu_i^x \rightarrow W_i= \prod_{\ell \in \square_i} \sigma_\ell^x$, while electric variables are represented by $\mu^z_i \mu_{i-1}^z \equiv \sigma_\ell^z$ where $\ell$ is the link between plaquettes $i$ and $i-1$. The corresponding ``bulk'' Hamiltonian terms for the subsystem $A$ then reads
\begin{align}\label{eq:HA}
H_{A}^{\rm dual} =  \sum_{{i} \in {A_{\rm bulk}}} \mu^x_{i}  + g\, \big(\sum_{\langle i,j \rangle \in A}
\mu^z_{i} \mu^z_j
+\kappa \sum_{i\in A} \mu^z_i\big),
\end{align}
where $\kappa \equiv 1+V_y$. Here, $i \in A_{\rm{bulk}}$ in the sum over $\mu_i^x$ indicates that the original plaquette $\square_i$ is entirely in $A$ (i.e., does not touch the boundaries). $\langle i,j \rangle\in A$ denotes nearest-neighbor bulk-spin pairs $i,j$ in $A$. Finally, the sum over $\mu_i^z$ runs over all bulk-spin indices $i$. The terms in the Hamiltonian acting on the complement are defined identically. Importantly, at the boundaries between the subsystems, $\ell_{\bar{A}A}$ and $\ell_{A\bar{A}}$ [see the orange circles in Fig.~\ref{fig:model}
(b)], the gauge-variant variables of the LGT, as described by~\Eq{eq:Z2Hamiltonian}, are retained. Denoting the Ising-spin index at the $A$ side of one of the boundaries to be $k$, the Hamiltonian terms coupling the subsystems at this boundary are
\begin{align}
    H^{\rm dual}_{\bar{A}A} \equiv  \mu^x_{
    k}\sigma^x_{\ell_{\bar{A}A}} + \mu^x_{
    k-1}\sigma^x_{\ell_{\bar{A}A}} + g\sigma^z_{\ell_{\bar{A}A}} \,,
\end{align}
with a similar definition for the other boundary. Note that on the plaquette $k$ containing one boundary link, $\mu_k^x\sigma^x_{\ell_{\bar{A}A}} \equiv W_k$, where $ \mu_k^x\equiv \prod_{\ell \in \tilde{\square}_k} \sigma_\ell^x$, with $\tilde{\square}_i$ referring to all but non-boundary links of plaquette $k$, and similarly for the plaquette $k-1$. The Gauss laws at the boundaries are not eliminated by the duality. The two Gauss laws in the dual model that are independent (one at each boundary) are
\begin{align}
G^{\rm dual}_{\bar{A}A} \equiv \mu^z_
k\sigma^z_{\ell_{\bar{A}A}} \mu^z_{
k-1}\,,
\end{align}
and similarly for the boundary at $\ell _{A\bar{A}}$.
The dual Hamiltonian of the model is the sum of the terms above:
\begin{align}
\label{eq:H-dual}
H^{\rm dual} =H_{A}^{\rm dual} +H_{\bar{A}}^{\rm dual} + H^{\rm dual}_{\bar{A}A}+ H^{\rm dual}_{A\bar{A}}\,.
\end{align}

This formulation ensures that all gauge-invariant variables of the LGT and its dual have the same expectation values. Importantly, by maintaining the Gauss laws at the boundaries, the entanglement properties of the dual formulation are identical to those of the LGT~\cite{mueller2022thermalization}. This is a subtle but crucial distinction from the standard Ising duality~\cite{wegner2015duality}, which does not preserve the entanglement structure.

\begin{figure}[t]
  \centering
	\includegraphics[width=0.25\textwidth]{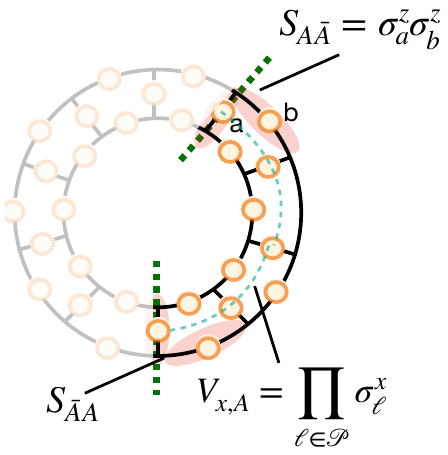}
	\caption{\textit{Symmetry operators.} 
Symmetry operators of the reduced density matrix of a state evolving under Gauss-law--respecting dynamics. The thin dashed green line represents the path $\mathcal{P}$ along which the Pauli operators in $V_{x,A}$ act on.
}
	\label{fig:symm}
\end{figure}

The entanglement structure depends on  the symmetries of the reduced density matrix, which stem from the remaining Gauss laws.
Specifically, $[S_j,\rho_A(t)] =0$,
where in LGT variables, $S_{j}\equiv \prod_{\ell_j \in A } \sigma^x_{\ell_j}$,
with $\ell_j$ being the two links originating from a boundary site with lattice coordinates $j=(j_x,j_y)$ within $A$. This is a direct consequence of the state being a Gauss-law eigenstate~\cite{mueller2022thermalization}. In addition, $[V_{x,A},\rho_A(t)] =0$ where the \emph{long-ribbon} operator within $A$ is defined as $V_{x,A}\equiv\prod_{\ell \in \mathcal{P}}\sigma^x_\ell$ where $\sigma^x_\ell$ act on the original degrees of freedom along the path $\mathcal{P}$ connecting both boundaries within $A$, see Fig.~\ref{fig:symm}.
Not all symmetry operators are independent.  For the quasi-1D chain of plaquettes shown in \Fig{fig:model}, only one of the two $S_j$ is independent at each boundary. We place one of the boundaries at sites $(0,0)$ and $(0,1)$ and the other at sites $(L_A,0)$ and $(L_A,1)$. We then choose the two independent $S_j$ operators to be $S_{(0,0)}$ and $S_{L_A,0}$, henceforth referred to as  $S_{A\bar{A}}$ and $S_{\bar{A}A}$, respectively. Note that, when acting on a Gauss-law--respecting state, $V_{x,A} =S_{A\bar{A}}S_{\bar{A}A}$. Consequently, there are four independent symmetry blocks $\rho_A(t)\equiv \bigoplus_{s=1,\cdots,4} \rho_{A,s}(t) $ that will play a role in our analysis.  In the dual formulation, these symmetry operators are simply $S_{A\bar{A}}^{\rm dual} \equiv \mu_k^z \sigma^z_{\ell_{\bar{A}A}}$ and $S_{A\bar{A}}^{\rm dual} \equiv \mu_{k'}^z \sigma^z_{\ell_{\bar{A}A}}$ where $k$ and $k'$ are the first and last plaquettes within $A$. We denote the eigenvalue $\{1,1\}$, $\{1,-1\}$, $\{-1,1\}$, and $\{-1,-1\}$ sectors of $\{S_{\bar{A}A}^{\rm dual},S_{A\bar{A}}^{\rm dual}\}$ operators with labels $s=1,2,3,4$, respectively. 

\subsection*{Experimental Setup}
Our experimental system consists of a linear chain of fifteen $^{171}$Yb$^{+}$ ions trapped along the $\mathbf{x}$ axis of a microfabricated ion trap ~\cite{osti_1237003} and spaced by approximately $3.7$ $\mu$m.  The $| \da \rangle$ and $| \ua \rangle$ spin states are respectively mapped to the first-order magnetic-field--insensitive hyperfine $| F=1, m_F=0\rangle$ and $| F=0, m_F=0\rangle$ states of the $^2S_{1/2}$ ground electronic manifold of the ions. {Here, we use the quantum-information convention for the spin states, i.e., $\ket{\uparrow} \equiv \ket{0}$ and $\ket{\downarrow} \equiv \ket{1}$.} The spin states are initialized in the $|\ua\rangle$ state by optical pumping and are measured by coupling to the excited ${}^2P_{1/2}$, $\ket{F=0,m_F=0}$ state using a 369-nm laser, wherein the presence (absence) of emitted photons differentiates between the bright state $| \da\rangle$ and the dark state $| \ua\rangle$ with an average $0.3\%$ infidelity, limited by off-resonant photon scattering.

The ions are individually addressed by an equispaced array of 1 $\mu$m-diameter, 355-nm, laser beams oriented perpendicular to the trap surface ($\mathbf{z}$ axis), together with a global 300 $\mu$m $\times$ 30 $\mu$m beam oriented along the $\mathbf{y}$~axis, which is parallel to the trap surface (see \Fig{fig:introoverview}(a) and Ref.~\cite{eigen2018universal}). These beams drive stimulated Raman transition via the ${}^2P_{1/2}$ and ${}^2P_{3/2}$ excited electronic states, where a photon is absorbed from the global beam and emitted into an individual beam to flip the qubit from $| \da \rangle$ to $| \ua \rangle$ and vice versa. All spin-manipulation light is produced by a single pulsed laser, modified to control its repetition rate to null 4-photon Stark shifts. The phase and amplitude of the 355-nm beams is controlled by single-channel and 32-channel acousto-optic modulators provided by L3 Harris Corporation.
\begin{figure*}[t]
  \centering
	\includegraphics[width=0.99\textwidth]{BW_Ansatz.pdf}
	\caption{\emph{Operator set for the entanglement-Hamiltonian ansatz.} Representation of the operator set in the entanglement-Hamiltonian ansatz used in our study, encompassing operators positioned inside the subsystem. All operators are Hermitian. Electric fields along the $x$ direction are interdependent (since $V_y=1$), thus not treated as independent operators. For the specific subsystem with 
 $N_A=4$, our ansatz incorporates a total of 73 distinct operators. While this is close to the subsystem Hilbert-space dimension, for a larger system, the number of operators scales linearly with the degrees of freedom within the subsystem.
 }
	\label{fig:BWansatzops}
\end{figure*}
\begin{figure}[t]
  \centering
	\includegraphics[width=0.32\textwidth]{BW_Ansatz_symmetries.pdf}
	\caption{\emph{The operator related to the symmetry sectors of the ansatz state.} Because of the (remaining) Gauss laws, some operators outside of the  subsystem are identical to operators within its complement. The operator shown, and a similar operator placed on the other boundary, commutes with all other operators in the BW-inspired ansatz shown in \Fig{fig:BWansatzops}. Therefore, their (common) eigenspace are the symmetry sectors of the reduced density matrix.
    }
	\label{fig:BWansatzopsbdry}
\end{figure}
To minimize addressing errors, crosstalk, and stray coupling to axial motion of the ions \cite{cetina2022cot}, single-qubit operations are realized using compound SK1 pulses \cite{brown2004aac} with Gaussian-shaped sub-segments, with typical infidelities of $0.2\%$. Entangling operations between any two qubits are realized via variable-angle pairwise M{\o}lmer-S{\o}rensen (MS) gates~\cite{sorensen1999quantum} employing laser-induced state-dependent forces on the $\approx 3$-MHz motional modes of the ion chain oriented along the $\mathbf{y}+\mathbf{z}$ axis. Robust decoupling from the ion motion at the end of each entangling gate is accomplished using amplitude-modulated, detuning-robust pulse waveforms \cite{Leung2018robust}. Typical non-unitary errors of fully-entangling MS gates are 1\% and consist predominantly of gate-angle errors and $Z$-flips of the individual qubits \cite{huang2023comparing}.
 \begin{figure*}[ht]
  \centering
	\includegraphics[width=0.86\textwidth]{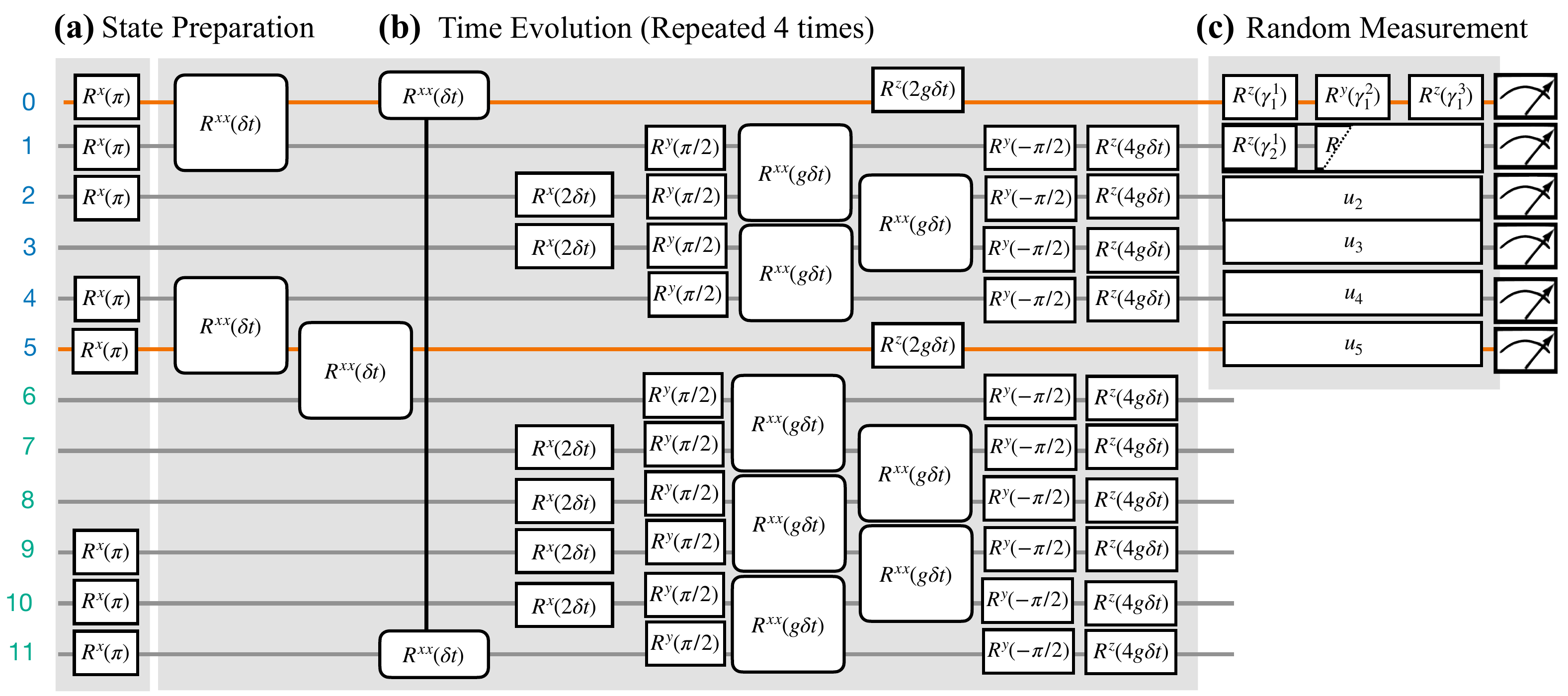}
	\caption{\textit{Overview of the circuits employed in the simulations of this work}. (a) Initial-state preparation, which involves setting a Gauss-law--respecting electric eigenstate in the Bloch $z$ basis. The shown circuit corresponds to one such initial state. (b) Trotterized time evolution, consisting of magnetic interactions (in the $x$ basis) comprised of single- and two-qubit rotations in the dual formulation. The electric part of the Hamiltonian evolution is diagonal consisting of $Z$ and $ZZ$ rotations. All MS gates are implemented according to \Eq{eq:Mssubst} depending on the employed time-evolution step $\delta t$. (c) Randomized-measurement circuits. $\gamma^{1,2,3}_i$ at qubit $i=0,1,\cdots,5$ are drawn from a circular unitary ensemble. The qubits $0$ and $5$ marking the two boundaries of the periodic lattice are drawn in orange. The qubit index of the rotation gates are dropped to reduce clutter and can be deduced from the qubit(s) they act on.
    }
	\label{fig:circuits}
\end{figure*}

\subsection*{Entanglement Hamiltonian Tomography}
Our randomized-measurement scheme applies one of $N_\mathcal{U}$ different gates $\mathcal{U}_j = \otimes_{i=0,\cdots,N_q-1} u_{i,j}$ for $j=1,\cdots,N_\mathcal{U}$ consisting of independent single-qubit gates $u_{i,j}$ sampled from a unitary Haar-design~\cite{dankert2009exact}. 
Concretely, randomization is via single-qubit random circuits, $\mathcal{U}\equiv \bigotimes_i u_i$ where $u_i$ is the following single-qubit unitary 
\begin{center}
\begin{adjustbox}{width=0.36\textwidth}
\begin{quantikz}
\gate{u_i {}^{}} & 
\end{quantikz}
=
\begin{quantikz}
\qw& \gate{R^z(\gamma^1_i)} & \gate{R^y(\gamma^2_i)}  & \gate{R^z(\gamma^3_i)}  & \qw
\end{quantikz}\,,
\end{adjustbox}
\end{center}
and for each qubit, the angles $\gamma^1_i$, $\gamma^2_i$, $\gamma^3_i$ are drawn according to a circular unitary ensemble (an overall phase is ignored) following Refs.~\cite{brydges2019probing,mueller2023quantum}. A drawback of this approach is that it does not maintain the symmetry structure of $\rho_A$, as it randomizes over the whole Hilbert space instead of each symmetry block of $\rho_A$. Such a symmetry-ignorant randomization mixes different symmetry sectors, resulting in an outcome that is inaccessible to any physical time-evolved quantum state. We choose it, nonetheless, to avoid the larger circuit depth associated with a symmetry-conscious randomization such as that proposed in Ref.~\cite{bringewatt2023randomized}. We note that a symmetry-conscious scheme would significantly reduce the measurement cost, as the sampling cost would only scale as the size of the symmetry block instead of the Hilbert-space size. It would thus simplify the classical optimization significantly. However, because the single-qubit scheme is tomographically complete, one can still reconstruct, approximately, the symmetry structure from the data.

The classical post-analysis  of the randomized measurement results consists of comparing the relative frequencies of the different bitstrings $b$ to the prediction based on an approximate EH inspired by the Bisognano-Wichmann (BW) theorem~\cite{bisognano1975duality,bisognano1976duality}. 
Explicitly, we assume that $H_{\rm E}(t)$ is a linear combination of $k-$local terms $\mathcal{O}_i$, i.e.
\begin{align}
        H_{\rm E} (t ; \{ \beta_i\} ) \equiv \sum_{i} \beta_i(t) \mathcal{O}_i\,.
\end{align}
To find a suitable set of operators $\{ O_i \}$, we proceed as follows. Starting from the operator content of the physical Hamiltonian in \Eq{eq:Z2Hamiltonian}, containing at most $4$-local terms ($2$-local in the dual formulation), new operators are generated iteratively by forming non-trivial commutators from the existing set. This process can be halted after two iterations (`commutators of commutators') resulting in a maximum of $7$-local operators within the LGT framework (or $3$-local in the dual formulation).

While the BW theorem hints at an optimal choice of operators $\mathcal{O}_i$ for ground states, our ansatz is heuristic, as a systematic ansatz for non-equilibrium states is not known. Our criteria for the operators are that i) they are local, involving operators with support on at most two neighboring plaquettes, ii) they are compatible with the symmetries of the reduced state $\rho_A$ (which will be discussed below), and iii) they are independent, meaning they cannot be transformed into each other through Gauss laws when acting on a physical state (although they are certainly not independent in the algebraic sense).

The operators we use are pictorially represented in~Fig.~\ref{fig:BWansatzops} plus all operators that are generated from those depicted upon translation. Here for the sake of generality, we represent them in terms of $Z_2$ variables, albeit, in practice, they are represented in the dual formulation in our algorithm. Figure~\ref{fig:BWansatzopsbdry} shows a special operator; this operator (and related examples) relates operators within the subsystem to operators in the complement via Gauss laws. Further, it commutes with $\rho_A$, and is hence connected to the symmetry blocks of $\rho_A$. All terms in Fig.~\ref{fig:BWansatzops} are obtained by considering the physical-Hamiltonian operators and by recursively commuting the physical-Hamiltonian operators, stopping at two recursions (i.e., commutators of commutators). Other selection strategies are also feasible.

Given the ansatz and following Ref.~\cite{kokail2021quantum}, we then minimize the functional
\begin{align}\label{eq:optimziaton}
    \left\langle \sum_{b}  \bigg[P_{\mathcal{U}}(b) - \text{Tr}\big[ {\mathcal{U}}^\dagger |b \rangle \langle b|  {\mathcal{U}} \rho_A(t ; \{ \beta_i\}) \big] \bigg]^2 \right\rangle_{\mathcal{U}}
\end{align}
where $P_\mathcal{U}(b)$ is the probability to measure a bitstring $b$ in the basis determined by $\mathcal{U}$ and
\begin{align}\rho_A(t ; \{ \beta_i\}) \equiv \frac{ e^{-H_{\rm E} (t ; \{ \beta_i\} )} }{ \text{Tr}[e^{-H_{\rm E} (t ; \{ \beta_i\} )}]}\,.
\end{align}
is the normalized reduced density matrix parameterized by $H_{\rm E}$.
Here, $\langle \cdot \rangle_{\mathcal{U}}$ is the average over random circuits.
Note that optimizing \Eq{eq:optimziaton} effectively weighs more favorably the largest Schmidt eigenvalues of $\rho_A$ because they, on average, contribute the most to any random observable. Because of this, the optimization more accurately reproduces the low-energy part of $H_{\rm E}$. 
The optimization is performed using \textsc{Matlab}'s~\cite{MATLAB:2010} non-linear \textsc{fmincon} optimization package with the `sqp' algorithm~\cite{fmincon-key}. Convergence and uniqueness of the obtained minimum have been cross checked for several data sets using \textsc{Matlab}'s \textsc{GlobalSearch} routine with default parameters~\cite{globalsearch-key}. All EH couplings are confined to the range $\beta_i \in [-50,50]$. We check explicitly that the routine does not come close to the boundary of the parameter regime. The EH that is obtained in this way is projected into symmetry sectors, which we then separately analyze. For this projection, the symmetry sector can be read off directly from the row and column numbers of the EH when they are interpreted as binary. The corresponding subblock $H_{\rm E,s}$ is then selected for further analysis, where $s$ labels one symmetry sector. For the EGRD analysis, we analyze the gap-ratio distribution separately for every sector, then combine the distributions and re-normalize the total distribution. For the ESFF analysis, we average over the sectors. Finally, EGRD and ESFF shown in the main text and appendices involve a combination/average over initial states. The error band and the error bars in the plots are the standard deviation for the symmetry-sector and initial-state averages. A systematic study of the performance of the EHT approach can be found in Supplementary Information sec.~3.

\subsection*{Circuit Representation for State Preparation, Evolution, and Measurement
\label{app:circuit}}

\noindent
In this Section, we provide circuits employed to experimentally realize state preparation, time evolution, and randomized measurement in the $Z_2$ LGT in (2+1)D within the dual Ising formulation.

Our basic circuit design is depicted in \Fig{fig:circuits}. We label qubits with indices $q=0,\dots,L-1$, with $L=12$. Qubit $q=0$ corresponds to the spin at $\ell_{\bar{A}A}$ positioned at the interface between subsystem $A$ and its complement, identifiable by an orange circle in \Fig{fig:model}
(b). 
Qubits $q=1,\dots,4$ represent the succeeding dual variables arranged counterclockwise at the center of plaquettes denoted by purple circles in \Fig{fig:model}
(b). Following this sequence, qubit $q=5$ resides at the opposite boundary marked as $\ell_{A\bar{A}}$, and so forth. 

Working in the electric eigenbasis and starting from an all-up spin state, we first perform single-qubit $\sigma^x$ bit-flip operations to select a randomly chosen product state in the $\sigma^z$ basis that is consistent with Gauss laws, i.e., $G_j \ket{\psi(0)}=\ket{\psi(0)}$ for all $j$. Recall that in the dual formulation, there are only two Gauss-law operators each located at one of the subsystem's boundaries: $G^{\rm dual}_{\bar{A}A} = \mu^z_{11}\sigma^z_0 \mu^z_1$ and $G^{\rm dual}_{A\bar{A}} = \mu^z_4\sigma^z_5 \mu^z_6$. Importantly, any spin configuration in either bulk, $A$ and $\bar{A}$, is physical, as in the dual formulation, only gauge-invariant degrees of freedom are kept in the bulk.

For time evolution, we utilize a first-order Trotter scheme with time-evolution operator
\begin{align}
\label{eq:Trotter}
U(t)\equiv e^{-itH^{\rm dual}}\equiv \prod_{\delta t} U(\delta t).
\end{align}
We take $\delta t$ to be a variable time step while keeping the Trotter depth fixed. In the simulations conducted in this work, $t/\delta t= 4$, resulting in states and observables closely resembling the exact time-evolved states at early times. However, for late times, Trotter effects become more pronounced.

$U(\delta t)$ in Eq.~\eqref{eq:Trotter} can be written as $U(\delta t)\equiv \prod_{a\in X,Z,XX,ZZ} U_a $ with $U_a \equiv  e^{-i\delta tH^{\rm dual}_a}$; $H^{\rm dual}_a$ are the respective 1- and 2-local operators of the dual Hamiltonian in Eq.~\eqref{eq:H-dual}, sorted in $X$ and $Z$ operations, as well as single- and two-qubit entangling gates. While $XX$, $X$, and $Z$ are native operations, $ZZ$ entangling operations are realized via basis transformation and usage of the native MS gate,
\begin{align}
    R^{zz}_{ij}(\alpha) \equiv R^{y}_i\left(-\frac{\pi}{2}\right) R^{y}_j\left(-\frac{\pi}{2}\right) \,  R^{xx}_{ij}(\alpha) \, R^{y}_i\left(\frac{\pi}{2}\right) R^{y}_j\left(\frac{\pi}{2}\right). 
\end{align}
Here, $R^{x/y/z}_i(\alpha) \equiv e^{-i\frac{\alpha}{2}\sigma_i^{x/y/z}} $ and $R^{xx}_{ij}(\alpha) \equiv e^{-i{\alpha}\sigma_i^{x}\sigma_j^{x}}$. The experimental errors of our MS gates increase with the absolute value of the gate angle. To minimize the gate error, all MS gates are optimized as follows. First, all angles are mapped to the regime $\alpha \in [-\pi,\pi]$. Within this range, we make the following substitution
\begin{align}\label{eq:Mssubst}
     R^{xx}_{ij}(\alpha) \rightarrow  \begin{cases}
         R^{xx}_{ij}(\alpha) & \text{if }|\alpha| \le \frac{\pi}{4}\\
         R^{xx}_{ij}(\alpha+\frac{\pi}{2}) R^{x}_i(\pi) R^{x}_j(\pi) & \text{if }|\alpha| \in (\frac{\pi}{4},\frac{3\pi}{4}]\\ & \,\&\, \,\alpha<0 \\
         R^{xx}_{ij}(\alpha-\frac{\pi}{2}) R^{x}_i(\pi) R^{x}_j(\pi) & \text{if }|\alpha| \in (\frac{\pi}{4},\frac{3\pi}{4}]\\& \,\&\, \,\alpha>0\\
          R^{xx}_{ij}(\alpha+\pi)  & \text{if }\alpha < -\frac{3\pi}{4}\\
          R^{xx}_{ij}(\alpha-\pi)  & \text{if }\alpha > \frac{3\pi}{4}
     \end{cases}
\end{align}
so that all MS operations are restricted to the range $|\alpha| \le \frac{\pi}{4}$. 

\section*{Data Availability}
\noindent
The data displayed in all figures are publicly accessible at  \href{https://gitlab.com/Niklas-Mueller1988/thermalization-z2-lgt}{this repository}.
The raw data generated in this study have been deposited in Duke University's Box cloud storage database. 
 The raw data are available under restricted access subject to intellectual property laws, non-disclosure agreements, and trade-secret protections, in accordance with all applicable laws, regulations, agreement terms and conditions, and the policies and directives of the authors' sponsors and affiliated institutions. Access can be obtained by contacting Marko Cetina, Zohreh Davoudi, or Niklas Mueller. 

\section*{Code Availability}
\noindent
The code and circuit designs presented in this manuscript are publicly accessible at  \href{https://gitlab.com/Niklas-Mueller1988/thermalization-z2-lgt}{this repository}. 

\bibliography{bibi.bib}

\section*{Acknowledgment}
\noindent
N.M. thanks H. Froland, A. Polkovnikov, M. Savage, M. Srednicki, X. Yao, T. Zache, P. Zoller, and the participants of the InQubator for Quantum Simulation (IQuS) workshop ``Thermalization, from Cold Atoms to Hot Quantum Chromodynamics ''(\url{https://iqus.uw.edu/events/iqus-workshop-thermalization/}) at the University of Washington in September 2023 for many valuable discussions leading to this work. M.C. thanks L. Feng for valuable discussions and help with the experimental setup leading to this work.
N.M. acknowledges funding by the Department of Energy (DOE), Office of Science, Office of Nuclear Physics, IQuS (\url{https://iqus.uw.edu}), via the program on Quantum Horizons: QIS Research and Innovation for Nuclear Science under Award DE-SC0020970. Z.D., M.C., and T.W. were supported by the National Science Foundation's Quantum Leap Challenge Institute for Robust Quantum Simulation under Award OMA-2120757. Z.D. further acknowledges support by the DOE, Office of Science, Early Career Award DE-SC0020271. This work is further supported by a collaboration between the US DOE and other Agencies. This material is based upon work supported by the DOE, Office of Science, National Quantum Information Science Research Centers, Quantum Systems Accelerator. 

\section*{Author Contributions}
\noindent
M.C., Z.D., O.K., and N.M
designed the research; N.M. performed  the numerical computations, designed the quantum circuits, and conducted the data analysis; M.C. and T.W. performed experiments and collected the data; M.C., Z.D.,  and N.M. contributed to the manuscript, with input from all authors.

\section*{Competing Interests}
\noindent
M.C. is a co-inventor on  patents that are licensed from the University of Maryland to IonQ, Inc. The remaining authors declare no competing interests.

\section*{Supplementary Information}
\setcounter{equation}{0}
\setcounter{figure}{0}
\setcounter{table}{0}
\setcounter{page}{1}
\setcounter{section}{0}
\makeatletter
\renewcommand{\theequation}{\arabic{equation}}
\renewcommand{\figurename}{Supplementary Figure}
\renewcommand{\thesection}{\arabic{section}}
\renewcommand{\bibnumfmt}[1]{[#1]}
\renewcommand{\citenumfont}[1]{#1}

\section{Gauge Invariance Violation}
\begin{figure}[t]
  \centering
	\includegraphics[width=0.4\textwidth]{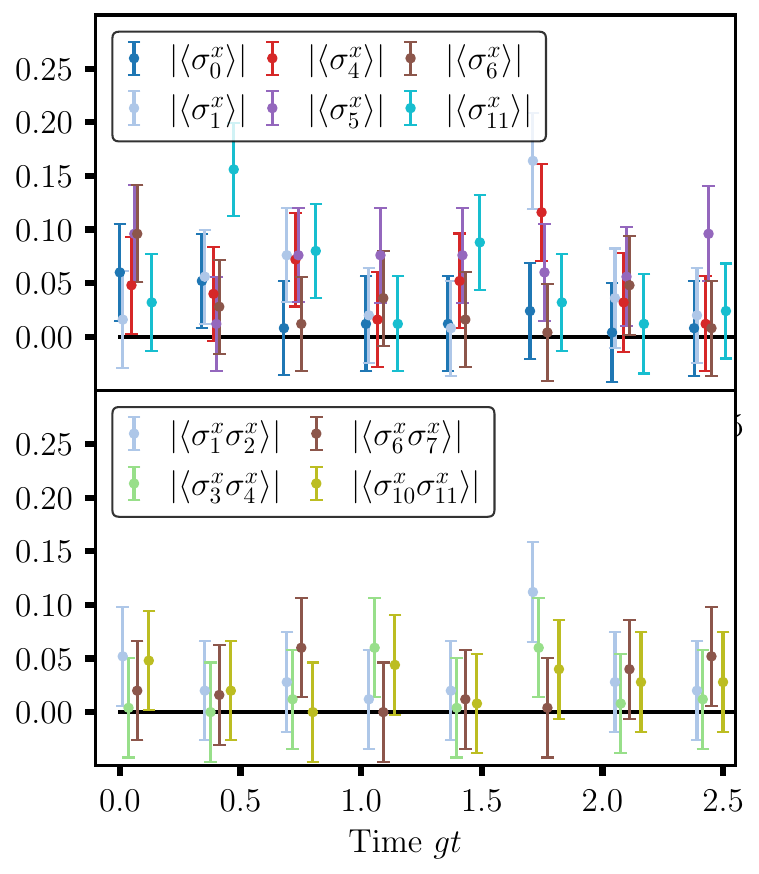}
	\caption{\textit{Estimation of gauge-invariance violation.} Expectation values of several single- and two-qubit non-gauge--invariant operators that are expected to be zero at all times. For better visibility, we plot the  different data sets slightly shifted in time $gt$.} 
	\label{fig:gauge_invariance}
\end{figure}

\noindent
An example study of the initial-state preparation, time evolution, and measurement in a fixed (Bloch $x$ or $z$) basis is shown in 3 (b) and (c) of the main text, where we presented measurement of several gauge-invariant and non-gauge--invariant
one- and two-qubit observables.  In addition, we present the measurement of several other non-gauge--invariant observables in \Fig{fig:gauge_invariance}. These observables are expected to be zero, but remain non-vanishing because of finite measurements and device errors. A main source of errors are likely coherent errors, related to the over- and under-rotation of single- and two-qubit gates. While the time-evolution circuit only contains gauge-invariant operations for any gate angle employed, the final single-qubit rotations $R^y_i(-\pi/2)$ and $R^x_i(\pi/2)$ (that transform all qubits $i$ from the Bloch $z$ basis into the Bloch $x$ and $y$ bases, respectively) can introduce gauge-invariance violation if their rotation angle is set inaccurately. Additionally, errors stemming from initial-state preparation and readout processes can contribute to this violation.

\begin{figure*}[t]
  \centering
	\includegraphics[width=0.9\textwidth]{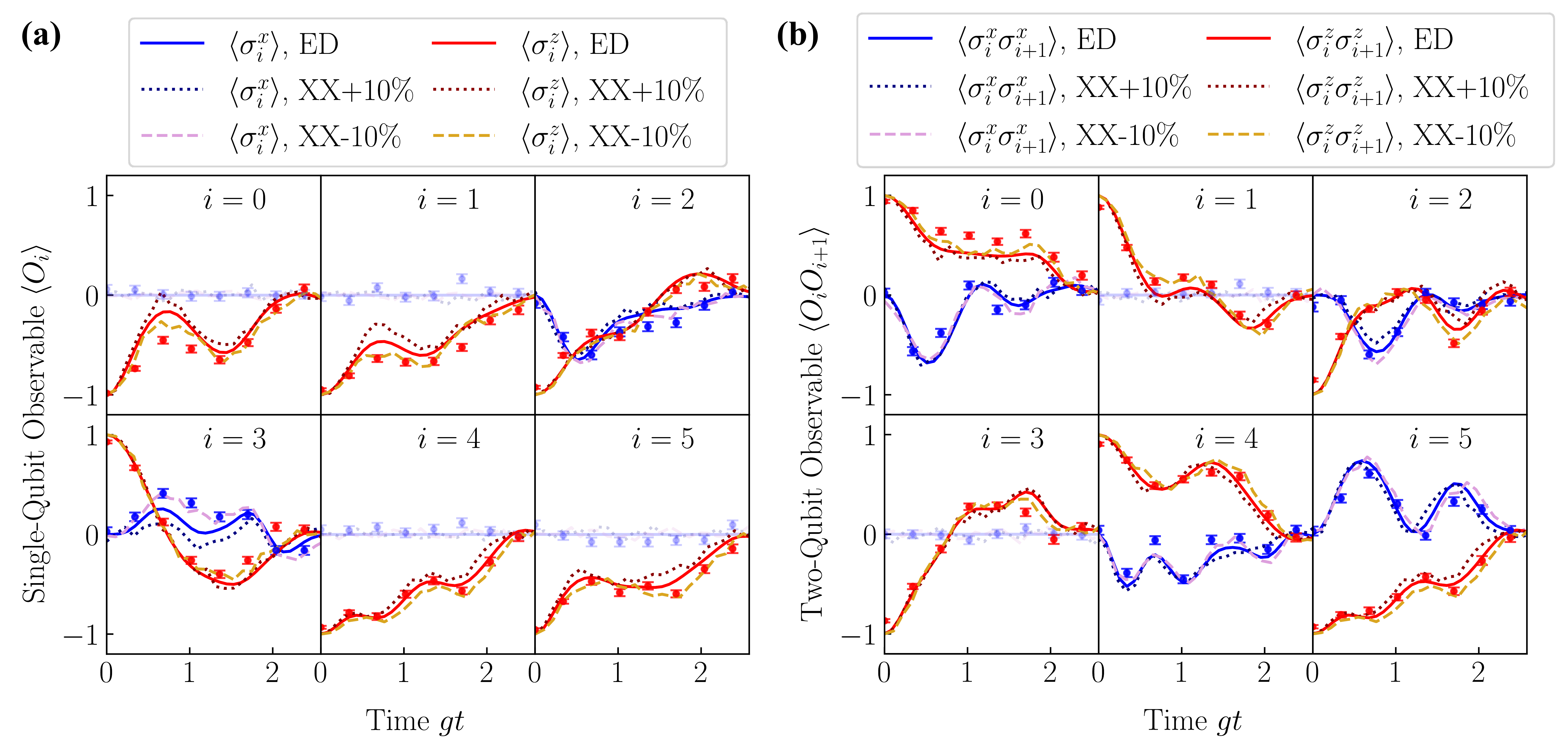}
	\caption{{\textit{Investigation of the sensitivity of local observables to systematic variation in XX gate angles.} Plots depict single-qubit (a) and two-qubit (b) observables as in Fig.~3 of the manuscript. Red and blue points denote experimental data from Fig.~3 of the manuscript; Red and blue solid lines are exact-diagonalization (ED) values from Fig. 3; Dark-blue and dark-red dotted lines are simulator data with all XX gate angles increased by 10\% (XX+10\%); Plum and  gold dashed lines are simulator data with all XX gate angles decreased by 10\% (XX-10\%).}
	\label{fig:fig3sensitivity}}
\end{figure*}

\begin{figure*}[t]
  \centering
	\includegraphics[width=0.9\textwidth]{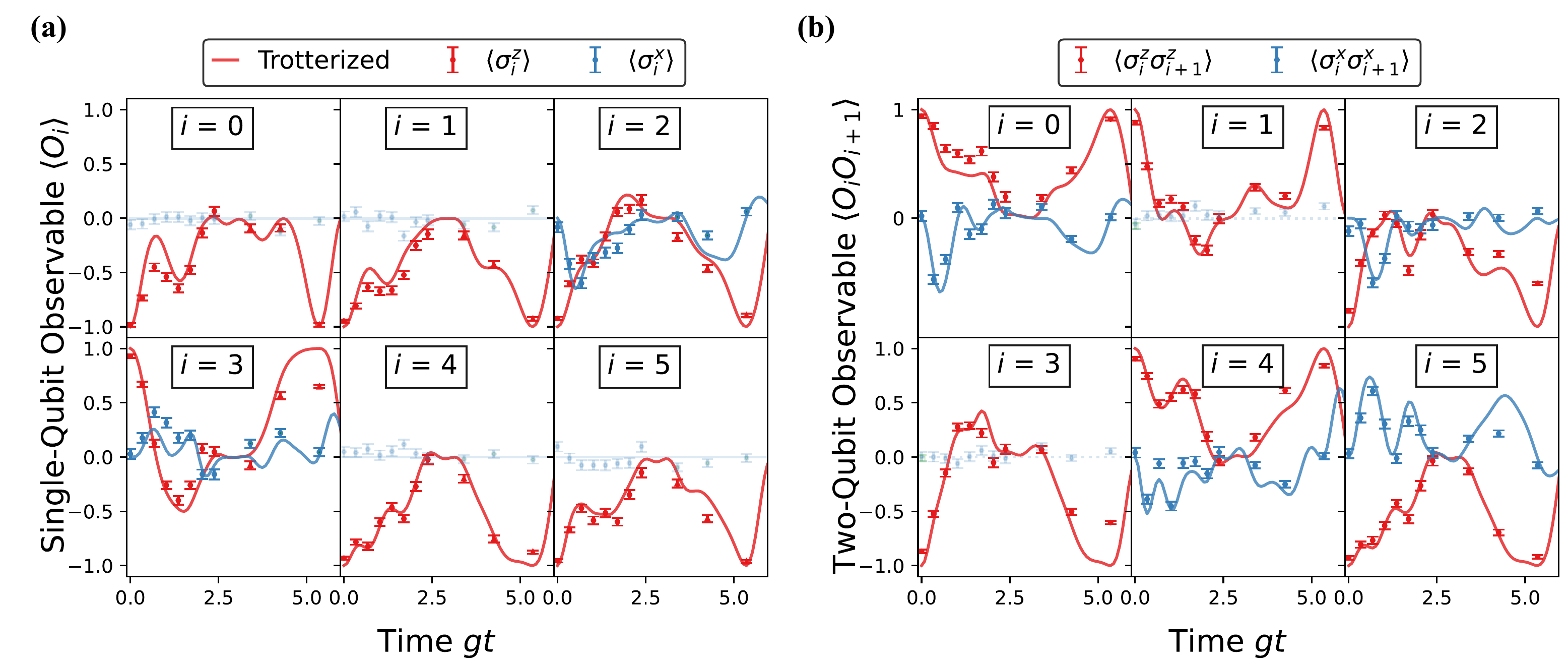}
	\caption{{\textit{Fixed-basis measurements at later times}. (a) Single-qubit observables in  $z$ basis (red) and $x$ basis (blue) including at later times as shown in the main text. (b) Two-qubit observables.  Dimmed out curves represent non-gauge-invariant observables. All data points are recorded with $t/\delta t =4$ Trotter steps. 
    Due to the coarse Trotterization, a recurrence of the initial state is observed, as indicated by the observables returning close to their initial values at later times.}}
	\label{fig:fixedbasis-late-time}
\end{figure*}
\begin{figure*}[t]
  \centering
	\includegraphics[width=0.93\textwidth]{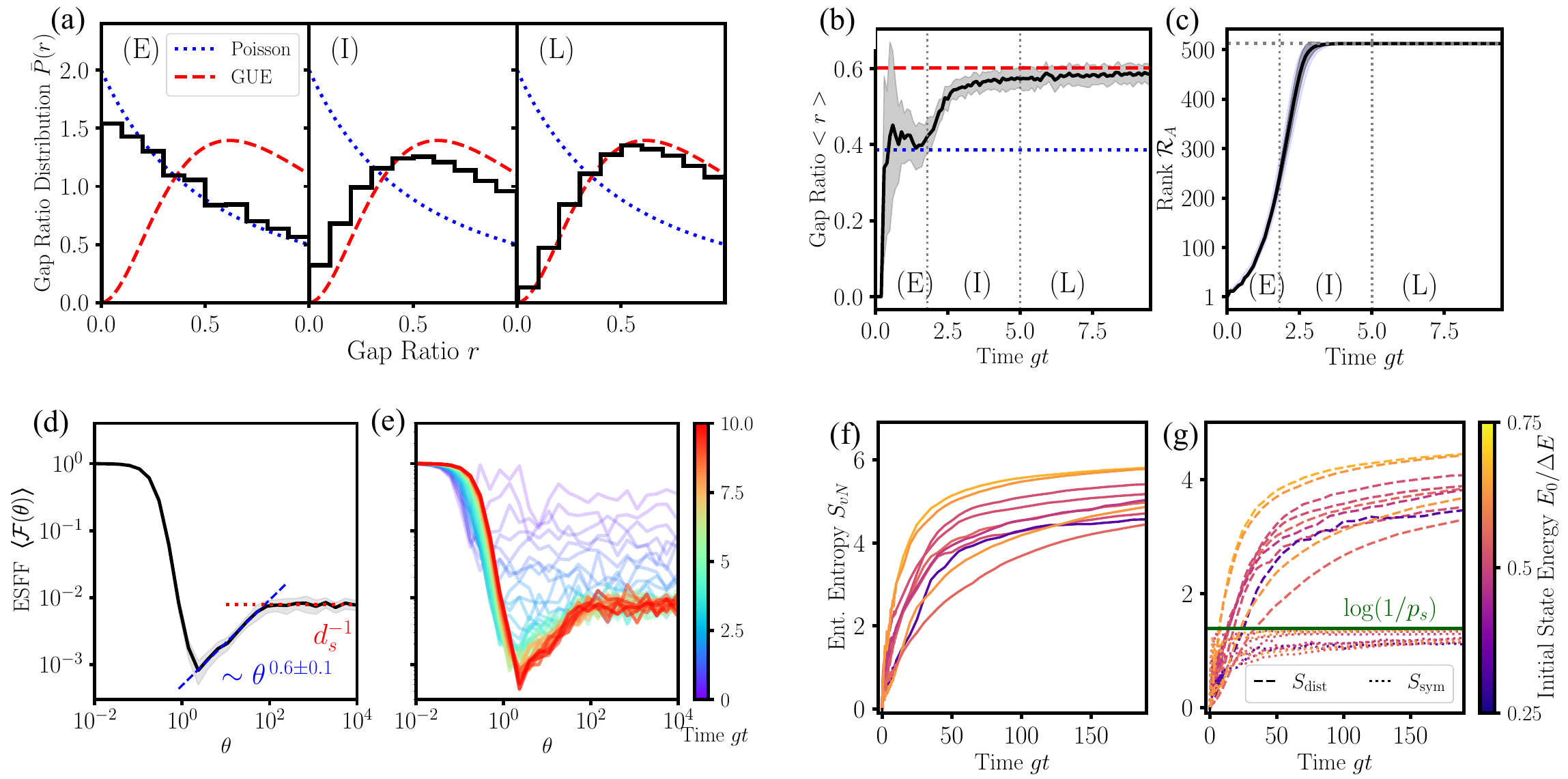}
	\caption{\textit{Theoretical analysis for $L=22$ and $L_A=10$ plaquettes.} (a) Entanglement-spectrum gap-ratio distribution, combining the gap ratios of ten randomly selected initial states and across all symmetry sectors of the reduced state. The distributions are additionally combined in time over all data in each of regimes (E), (I), and (L), and binned over intervals $\Delta r =1/12$. (b) Mean of the gap ratio distribution, combining initial states and symmetry sectors.  (c) Rank of the reduced density matrix as a function of scaled time. (d) Spectral form factor of the EH, averaged over symmetry sectors and over 10 randomly selected initial product states, averaged over time range (L) ($5< g t \le 10 $).  (e) Time dependence of the entanglement spectral form factor. (f) The von Neumann entanglement entropy as a function of scaled time, for  10 randomly chosen initial product states. The color encoding represents the energy of the initial state relative to the energy bandwidth $\Delta E$ of the physical Hamiltonian. (g) Distillable versus symmetry components of the von Neumann entanglement entropy.}
	\label{fig:gap-ratio-theory-analysis}
\end{figure*}

\begin{figure}[t]
  \centering
	\includegraphics[width=0.45\textwidth]{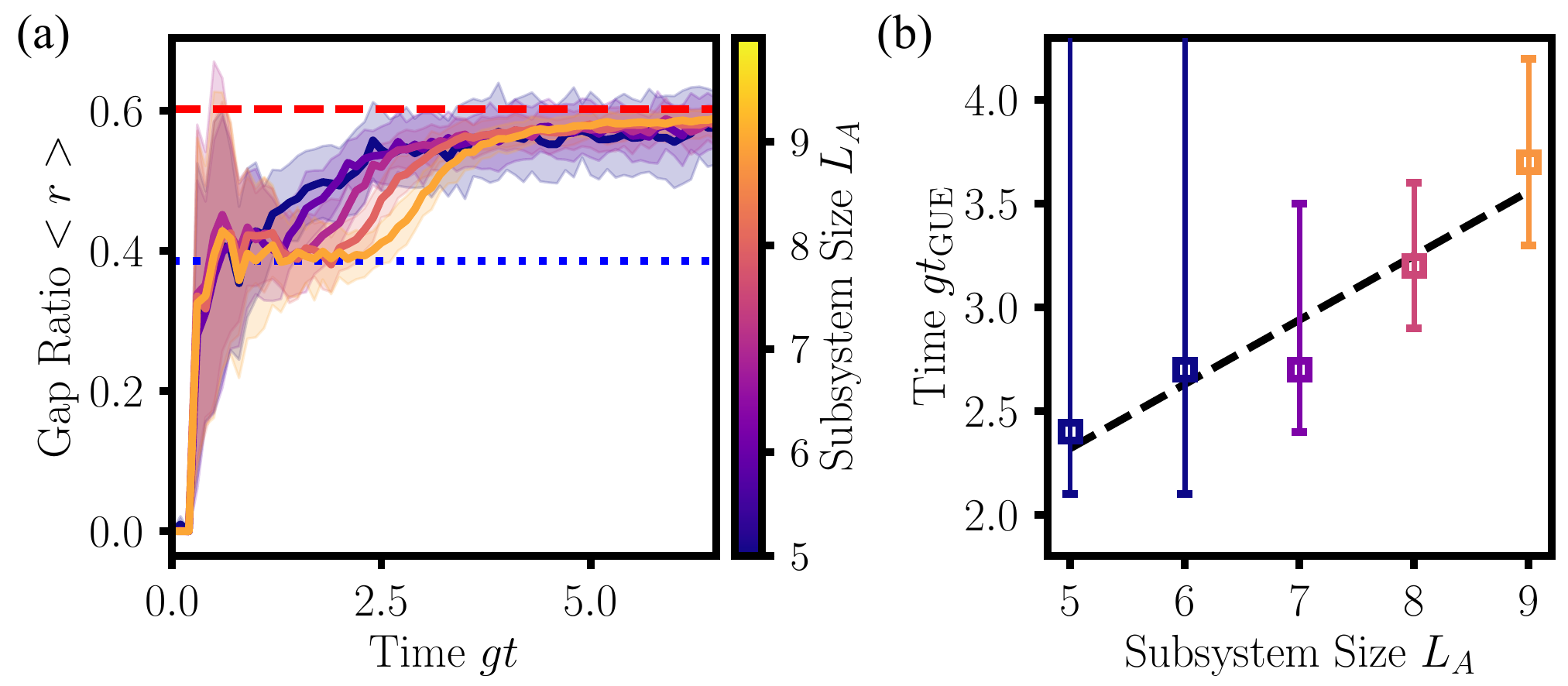}
	\caption{{\textit{Probing the subsystem-size dependence of the GUE time scale}. (a) Gap-ratio average of a numerically computed large system consisting of $L=22$ plaquettes evolving under exact time evolution while varying the subsystem size $L_A$. (b) Dependence of the time scale to reach GUE versus subsystem size, with uncertainties according to the description in the text. A dashed black lines indicates a linear fit to the central values; the uncertainty bands are not incorporated in the fit.}}
	\label{fig:gap_ratio_subsystem}
\end{figure}

\begin{figure}[h!]
  \centering
	\includegraphics[width=0.36\textwidth]{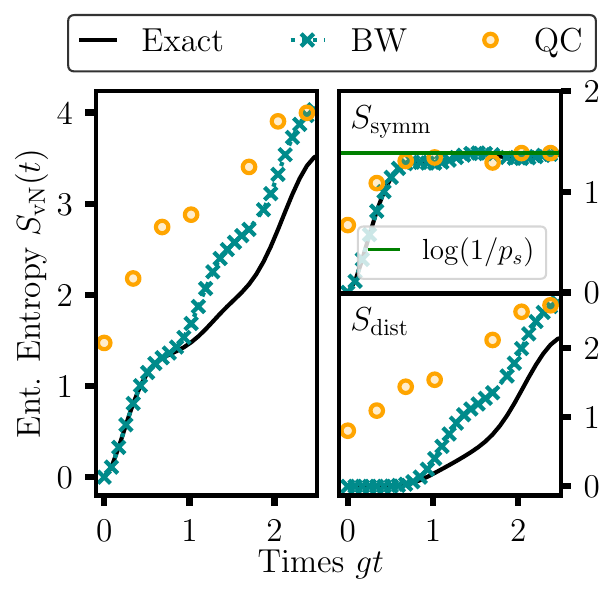}
	\caption{\textit{Entanglement entropy from experiment.} Shown is the von Neumann entanglement entropy, for one given initial state $\ket{\Psi_0}=\ket{\downarrow\downarrow\downarrow\uparrow\downarrow\downarrow\uparrow\uparrow\uparrow\downarrow\uparrow\uparrow}$,  in the left panel, comparing exact (black lines), infinite-measurement results using a BW-inspired parametrization (cyan crosses), and quantum-computed data (orange circles). The top and bottom side panels show the symmetry and distillable components of the von Neumann entanglement entropy, respectively. }
	\label{fig:vonNeumanExpt}
\end{figure}

{\section{Experimental Systematics}}

{The data in Fig. 3 show systematic deviations from the  theory predictions for trotterized evolution, hinting at an error in our quantum-computed evolution. Given the robustness of our SK1-based single-qubit gates, the most likely cause for this is a systematic error in the XX entangling gates. 

We investigated the sensitivity of our theory predictions in Fig. 3 to systematic variations in XX gate angles, and show the results of our analysis in \Fig{fig:fig3sensitivity}. Our simulations reveal that most of the observed deviations can be explained by a systematic $\sim$ 10\% \emph{under-rotation} in the XX gates. This under-rotation is consistent with a $\sim$150 nm drift in the alignment of the individual-addressing beams relative to the ions. This drift is consistent with recently measured beam drift between typical calibration runs on the same machine.}

\section{Finite-size Dependence and Late-time Behavior
\label{app:expanalysis}}
\noindent
In this section, we perform various tests aiming to extend our study toward larger systems, and to provide an outlook on the expected late-time behavior of non-equilibrium states, features that are inaccessible in current quantum computers. We present a detailed \textit{classical  computation} of a fairly large system ($L=22$, $L_A=8$ with coupling $g=1$), evolving the state continuously, and computing its entanglement structure using exact diagonalization, hence avoiding a Trotterization error. Such an analysis is possible for the model under consideration, owing to its relative simplicity. This simplicity enables us to achieve results that are fairly insensitive to finite-size effects, even with moderately large systems.  A similar classical computation will be impossible for more involved models or in higher dimensions.

{\Figure{fig:fixedbasis-late-time} presents the same fixed-basis single-qubit and two-qubit observables shown in the main text, in panels (a) and (b), respectively, with additional data points at later times. These data illustrate the effects of Trotterization: for a fixed Trotter step $ t/\delta t = 4 $, the observables return close to their initial values at late times, indicating a recurrence of the initial state. The experimental results remain consistent with exact numerical predictions (solid lines) even at these later times.}

We compute the same entanglement-related observables indicative of quantum chaos as in the main text, starting from several randomly drawn initial product states. The EGRD, i.e., the (normalized) distribution of the gap ratios $\bar{P}(r)$ with $r$ defined after Eq.~4 of the main text, is computed similarly as in the main text: we consider ten randomly chosen initial states, and combine the gap ratios that are separately computed for each symmetry sector of the respective reduced states and for all initial states. Additionally, we combine the distribution of gap ratios in three time regimes Early (E), Intermediate (I), and Late (L). The result is shown \Fig{fig:gap-ratio-theory-analysis}(a). Further, the mean of the gap ratio, combining initial states and symmetry sectors but not times, $\langle r \rangle$, is shown in \Fig{fig:gap-ratio-theory-analysis}(b) as a function of scaled time $gt$. 
The distribution in \Fig{fig:gap-ratio-theory-analysis}(b) is peaked near zero first, closely resembling a Poisson distribution. In this regime, the rank $\mathcal{R}_A\ll {d_A}$, where $ {d_A}$ is the Hilbert-space dimension of the subsystem. The effective rank $\mathcal{R}_A$ is defined after Eq.~5 of the main text and plotted for the present example in \Fig{fig:gap-ratio-theory-analysis}(c) starting from a product state with $\mathcal{R}_A=1$ at $t=0$. The largest $\xi_\lambda$ values in this regime correspond to extremely small probabilities $p_\lambda = \exp(-\xi_\lambda)$, close to or at the level of machine precision. A regularization of the smallest $p_\lambda$ is required: we manually cut off probabilities below $10^{-15}$ before computing $\mathcal{R}_A$, varying this limit by two orders of magnitude in each direction to provide the blue bands in panel \Fig{fig:gap-ratio-theory-analysis}(c), then using only the $\mathcal{R}_A$ lowest levels in the analysis of the EGRD and the ESFF. Gray bands represent the variance with respect to the randomly chosen initial states. $\mathcal{R}_A$
{This is necessary because both statistical measures, the EGRD and the ESFF, are typically used to describe the properties of \textit{physical} Hamiltonians, whose rank equals the Hilbert-space dimension. For EHs, we employ regularization suitable for non-full--rank matrices. However, experimental constraints, such as finite statistics and device errors, typically lead to EHs that are reconstructed as nearly or exactly full rank, rendering any regularization obsolete. Nonetheless, non-full--rank EHs occur for states computed in exact diagonalization, particularly at early times, that we contrast our data with.}

{Regime (I) marks a transitional phase where $\langle r \rangle $ grows further towards a Gaussian Unitary Ensemble (GUE), indicating quantum chaotic behavior.} At the end of this stage, $\mathcal{R}_A= {d_A}$ is maximal, and all probabilities $p_\lambda$ are well above machine precision, eliminating the need for any regularization. Finally,  regime (L) sees the saturation of the gap ratio to GUE level statistics (up to minimal finite lattice-size effects). The beginning and end of the time regimes (E-L) are chosen as follows, (E): $ 0 \le gt <1.8$, (I): $ 1.8 \le gt <5.0$, and (L): $ 5.0 \le gt<10.0$.
{Figure~\ref{fig:gap_ratio_subsystem} illustrates how the gap-ratio distribution approaches a GUE distribution as a function of subsystem size.}

An analogous picture is evident in the time evolution of the ESFF defined in Eq.~5 of the main text, and shown in Figs.~\ref{fig:gap-ratio-theory-analysis} (d) and (e). Panel (e) shows the ESFF for various times $0 \le gt \le 10$, starting from a flat distribution at earliest time and showing a plateau-ramp structure at late times. Similarly as in the main text, the ESFF is computed separately for every symmetry sector of the EH to avoid contamination from uncorrelated levels in different sectors, and then averaged over these sectors and over the randomly chosen initial states. Panel (d) shows the time average over regime (III). In this regime, the ramp-plateau is evident. A gray band marks the statistical deviation from the initial-state and symmetry-sector averages. We fit the ramp to the form $\sim \theta^{0.6\pm 0.1} $ which is consistent with the experimental data shown in the main text for a much smaller system. The fit error is determined by varying the fit range.

Because the late-time behavior, i.e., $g t\gg 1$, is inaccessible to current Trotter-based digital quantum simulation, we extend our analysis towards this regime, aiming to elucidate the potential outcomes that future quantum simulators might uncover. A similar analysis has been performed previously in Ref.~\cite{mueller2022thermalization}. At late times, it is expected that the entanglement entropy of a subsystem becomes equal to the thermal entropy contained in that subsystem, corresponding to a global Gibbs state with a temperature corresponding to the average energy of the initial state. In \Fig{fig:gap-ratio-theory-analysis} (f), we plot the von Neumann entanglement entropy, for a variety of randomly chosen initial states. The individual curves are color coded relative to their initial average energy density $\sim E_0 = \langle\psi(0)| H | \psi(0) \rangle$, normalized relative to the energy bandwith $\Delta E$,  i.e., the difference between the highest and lowest eigenvalues of the EH. Displayed are only initial states whose energies lie within the interquartile range, representing the central 50\% energy spectrum of states. States highlighted in yellow denote the highest energies, while those in blue represent the lowest. We were not able to extend this study to even later times to observe the expected saturation of the von Neumann entanglement entropy, owing to the fact that eventually finite-size effects become large. These results demonstrate that entanglement-entropy saturation is not a practical measure of thermalization in present experiments, demanding long evolution times and exhibiting lack of initial-state insensitivity (i.e., universality).

Shown in \Fig{fig:gap-ratio-theory-analysis}(g) is a separation of the von Neumann entanglement entropy, $S_{vN}$, into components, $S_{vN,s}$, related to the symmetry structure of  $\rho_A = \bigoplus_s \rho_{A,s}$,
\begin{align}
    S_{vN} = -\sum_s p_s \log(p_s) + \sum_s p_s S_{vN,s},
\end{align}
where the first term is the symmetry component, and the second is the distillable entanglement~\cite{ghosh2015entanglement,van2016entanglement}. Here, $p_s\equiv \text{Tr}[\rho_{A,s}]\le 1$ with $\sum_s p_s=1$, and $S_{vN,s} \equiv - \text{Tr}[\bar{\rho}_{A,s} \log(\bar{\rho}_{A,s})]$ is the sector-wise entanglement entropy with $\bar{\rho}_{A,s}\equiv \rho_{A,s} / p_s$. Dashed lines in the plot denote the distillable component, while dotted lines denote the symmetry part. It is evident that both components saturate on different time scales: the symmetry component saturates fairly quickly (albeit still later than the build-up of level repulsion) to its maximal value $\log(1/p_s)$ where $p_s =1/4$ corresponds to equal mixing of the four symmetry sectors of $\rho_A$. In contrast, the distillable components dominates the late-time behavior of the von Neumann entropy.

{\Fig{fig:gap_ratio_subsystem} shows the subsystem size dependence of the saturation time where the average of the gap-ratio distribution is consistent with GUE predictions to within 90\%. A linear fit to the central values is shown in pnael (b), the uncertainty bands from panel (a) are not included in the fit.}

While experimental constraints prevent us from accessing late times, \Fig{fig:vonNeumanExpt} illustrates the von Neumann entanglement entropy derived from our experimental data, utilizing the same dataset as in the main text, for one given initial state $\ket{\Psi_0}=\ket{\downarrow\downarrow\downarrow\uparrow\downarrow\downarrow\uparrow\uparrow\uparrow\downarrow\uparrow\uparrow}$. In this figure, exact classically computed results are denoted by black lines, while cyan crosses represent the optimal BW-inspired parameterization assuming infinite measurements, and orange circles are the experimental data points. The right-hand side of the figure displays two panels: the top panel depicts the symmetry component, while the bottom panel depicts the distillable component. Notably, the observed behavior closely mirrors the (classically computed) findings of the much larger system in \Fig{fig:gap-ratio-theory-analysis}(g). Specifically, the symmetry component saturates at its maximum value of $\log(4)$, while the distillable entanglement demonstrates continued growth.  Generally,  the entanglement measured in our experiment via BW-inspired tomography overshoots the exact result. We attribute this discrepancy primarily to  over- or under-rotations within the single-qubit random-bases changes. 

\begin{figure}[t]
  \centering
	\includegraphics[width=0.35\textwidth]{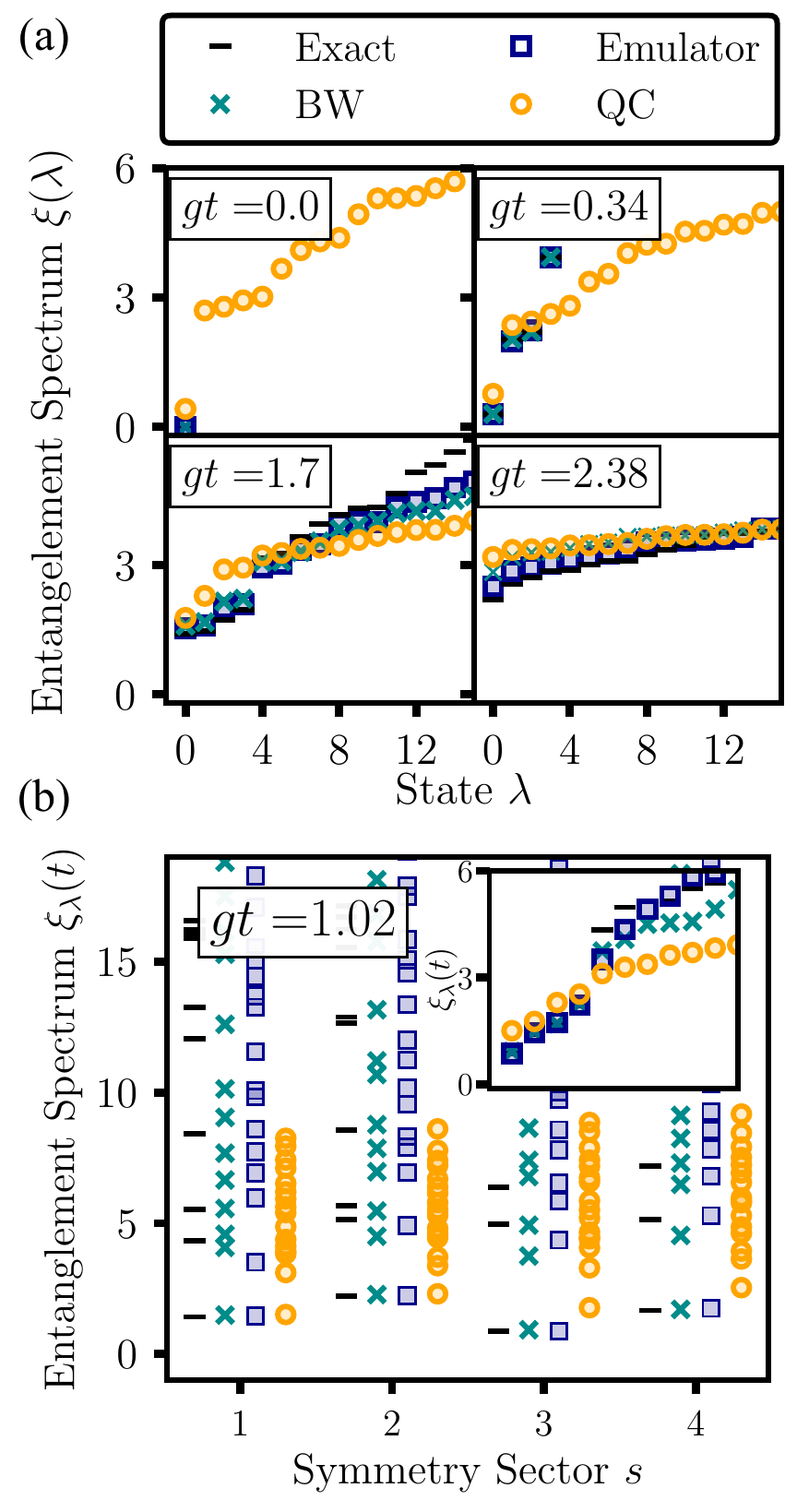}
\caption{\textit{Entanglement spectrum from tomography.} (a) Lower parts of the entanglement spectrum for times $gt=$ 0, 0.34, 1.7, and 2.38, comparing exact data (black lines), infinite-measurement results (cyan crosses), emulator data (blue squares), and quantum-computed data (orange circles), for one given initial state $\ket{\Psi_0}=\ket{\downarrow\downarrow\downarrow\uparrow\downarrow\downarrow\uparrow\uparrow\uparrow\downarrow\uparrow\uparrow}$. (b) Entanglement spectrum separated into the four symmetry sectors for the same data set at $gt$=1.02. The inset shows a close-up of the lower part of the spectrum for all symmetry sectors.
}
	\label{fig:overviewES}
\end{figure}

\begin{figure}[t]
  \centering
	\includegraphics[width=0.3\textwidth]{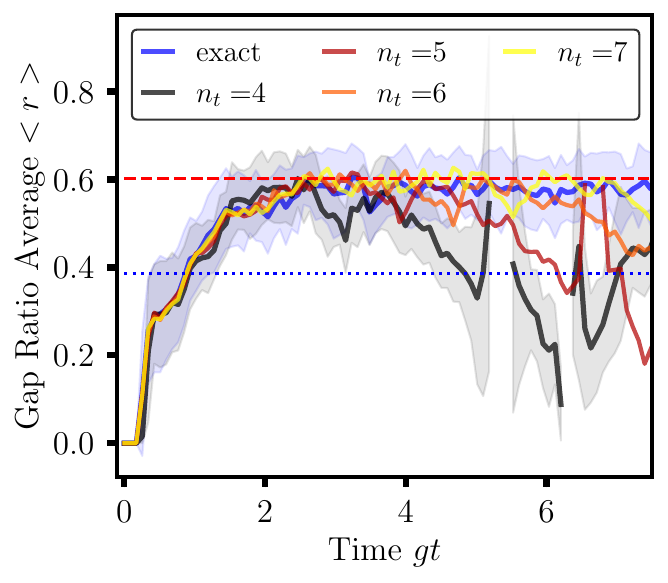}
\caption{{\textit{Trotterization.} Trotterization with coarse step size $n_t \coloneq t /\Delta t$, when compared to exact time evolution (blue solid lines), results in recurrences of the initial state. The numerically computed average gap ratio is shown, which initially approaches a value consistent with the Gaussian Unitary Ensemble (GUE), but subsequently decreases, eventually becoming 'un-chaotic' due to the recurrence of the initial state. Gaps in the curves correspond to states of such low rank preventing the determination of the gap-ratio distribution.}}
	\label{fig:Trotter}
\end{figure}

\section{Performance of the Entanglement-Hamiltonian Tomography Protocol
\label{app:EHTanalysis}}
\noindent
In this section, we systematically investigate the performance of the approach with regard to the number of bases sampled, the number of shots performed, and the influence of device errors. {In addition, we study the effects of Trotterization.}
\begin{figure}[t!]
  \centering
	\includegraphics[width=0.45\textwidth]{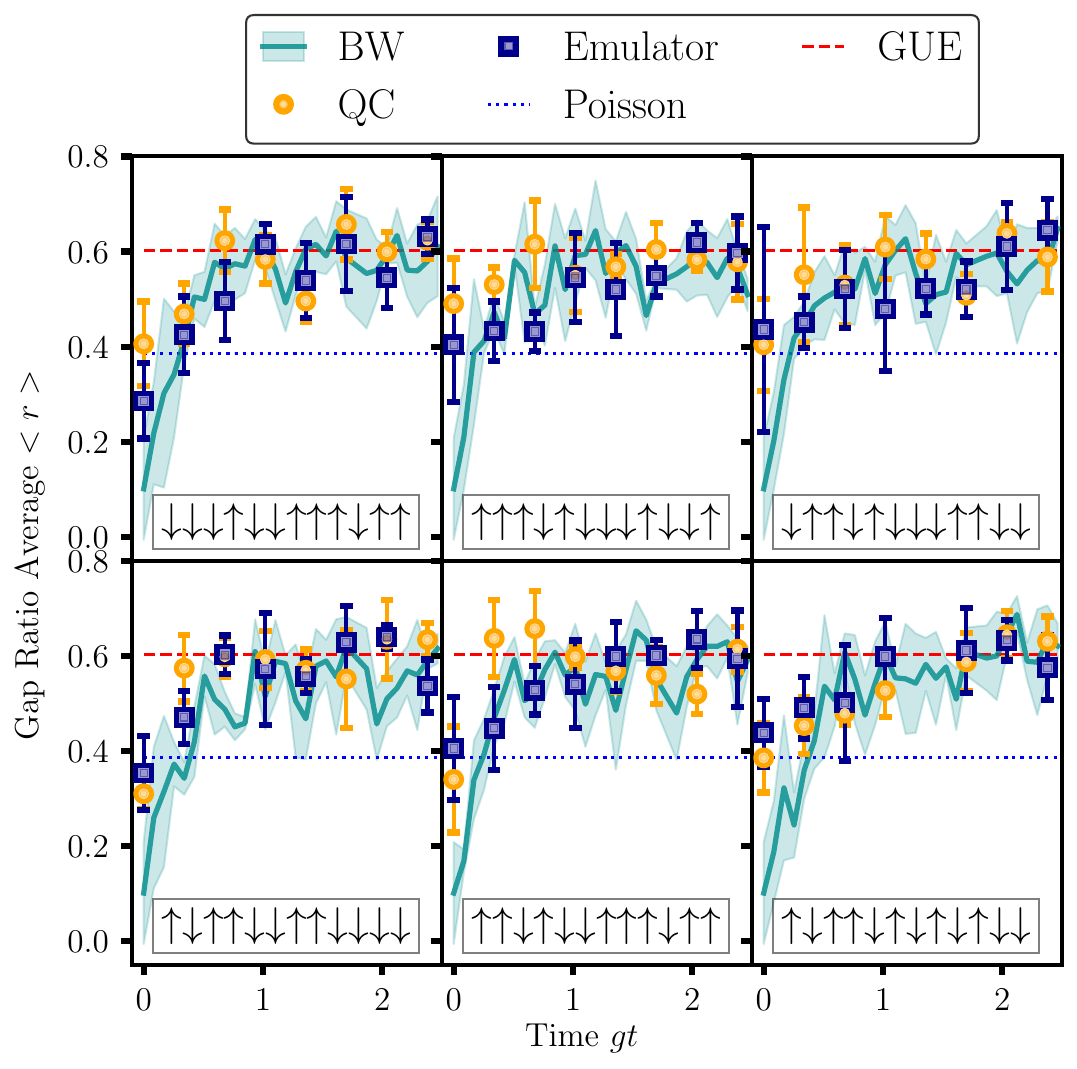}
	\caption{\emph{Gap-ratio averages across initial states.} Comparison of the average gap ratio obtained in experiment (yellow circles) versus using an ideal emulator (blue squares), shown individually for six randomly-drawn initial electric eigenstates compatible with the Gauss laws. We show the spin configuration of the initial state in each panel, which is randomly drawn. Error bars and error bands represent the spread over symmetry sectors. Dotted blue and dashed red lines denote the values associated with the Poisson distribution and GUE, respectively.}
	\label{fig:BWgapratiostates}
\end{figure}

\begin{figure}[t!]
  \centering
	\includegraphics[width=0.43\textwidth]{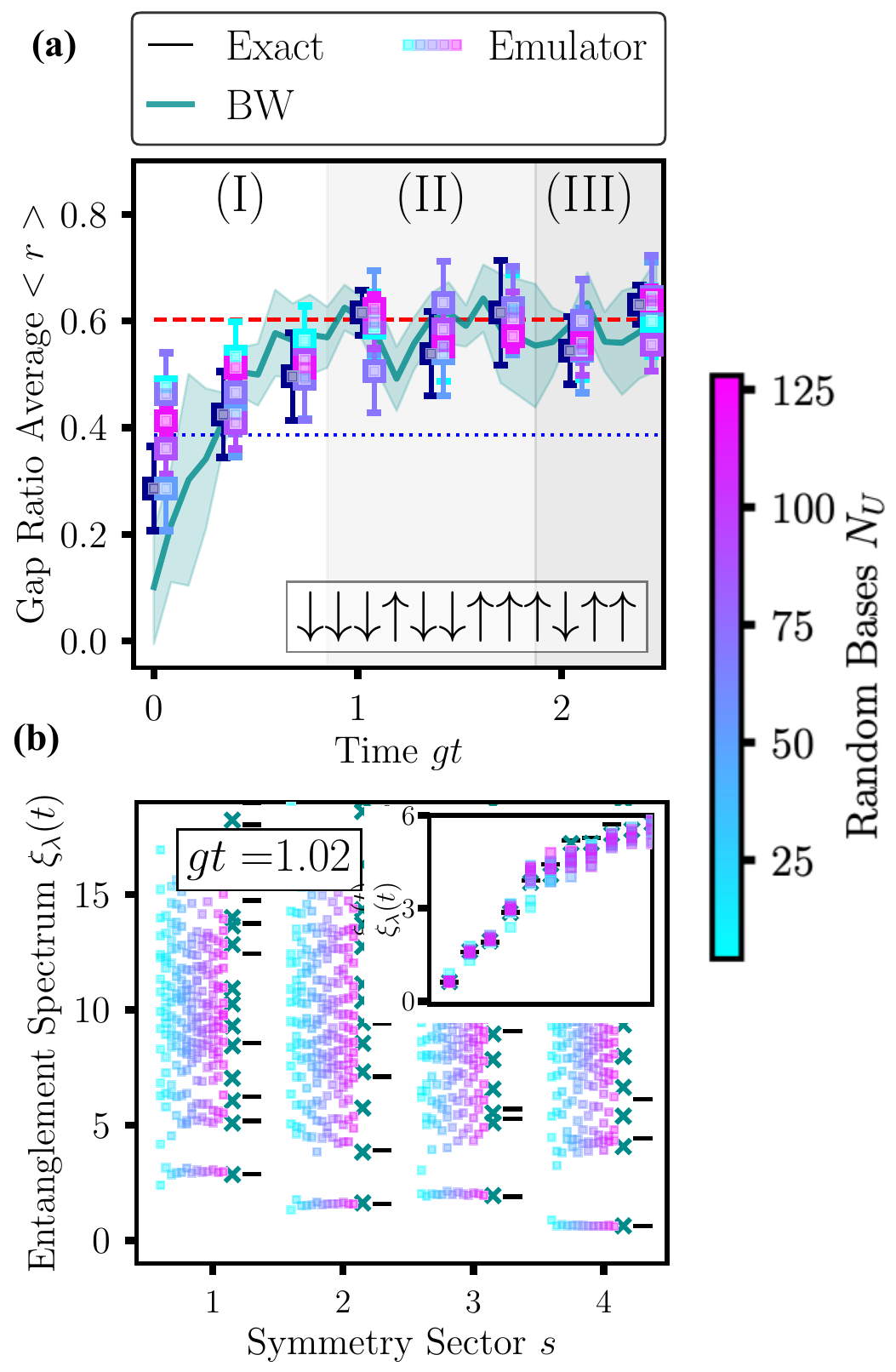}
\caption{\textit{Estimation of basis dependence (emulator data).} (a) Estimation of dependence on the number of random bases $N_\mathcal{U}$, illustrated for fixed $N_{\rm shots}=750$ per basis for one representative initial state. Error bars and error bands represent the spread over symmetry sectors. (b) Example of a reconstructed entanglement spectrum, varying the number of random bases for a fixed number of shots $N_{\rm shots}=750$. The inset shows a close-up of the lower part of the spectrum for all symmetry sectors.}
	\label{fig:BWgapratiobases}
\end{figure}
\begin{figure}[t!]
  \centering
	\includegraphics[width=0.43\textwidth]{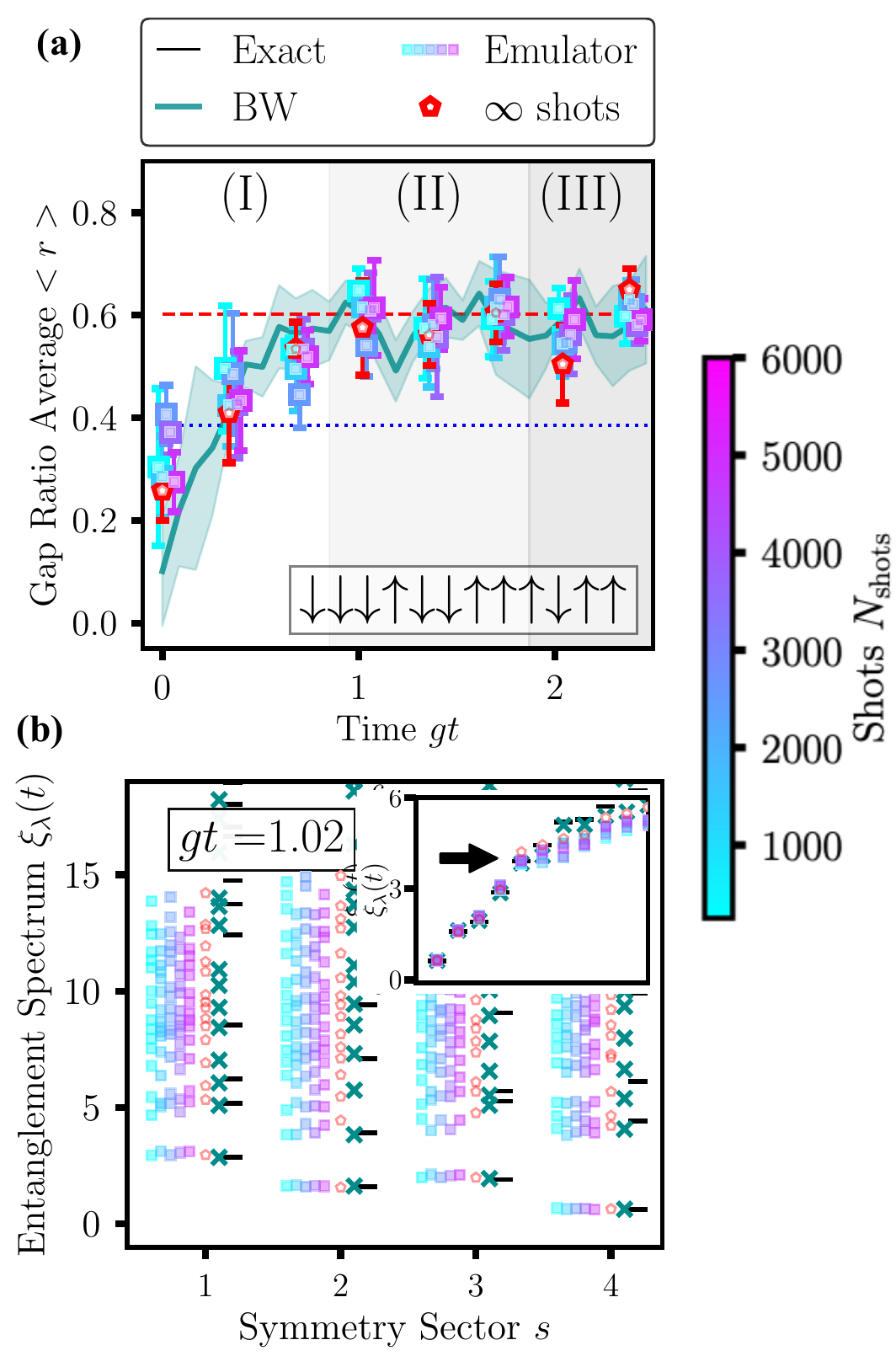}
\caption{\textit{Estimation of shot-noise dependence (emulator data).} (a) Estimation of dependence on the number of shots, illustrated for fixed number of random bases $N_{U}=24$ for one representative initial state. Error bars and error bands represent the spread over symmetry sectors. (b) Example of a reconstructed entanglement spectrum, varying the number of shots for a fixed $N_{U}=24$. Red hexagons represent the infinite-shot limit. The inset shows a close-up of the lower part of the spectrum for all symmetry sectors, a black arrow indicates where emulator data (in the infinite-shot limit) diverges from the ideal BW results.}
	\label{fig:BWgapratioshots}
\end{figure}

An example of the eigenvalue spectrum of the EH, comparing the exact values with those obtained from our procedure, is shown in~\Fig{fig:overviewES}. Panel (a) shows the reconstructed entanglement spectrum for several times, comparing exact data (black lines), infinite-measurement results (cyan crosses), emulator data (blue squares), and quantum-computed data (orange circles), for one given initial state. Panel (b) shows the symmetry-resolved spectrum for $gt$=1.02. Here, $s$  are the symmetry sectors of $\rho_A$, see the discussion in the Methods section. While our analysis effectively describes the low-lying part of the entanglement spectrum, it fails to quantitatively reproduce the higher part associated with very small probabilities. Additionally, the reconstructed spectrum appears more mixed (more entangled) compared to e.g., the emulator data. This discrepancy primarily is due to finite-measurement statistics and device errors. Comparison with the infinite-(ideal-)measurement results also reveals that the BW-inspired parameterization falls short in capturing  quantitatively higher-lying components of the spectrum. It is noteworthy that the limitations of the parametrization become more pronounced at later times where the Trotter-step size is large and even the low-lying part deviates. This is because the effective Trotter Hamiltonian at late times deviates from the known target Hamiltonian from which the BW-inspired ansatz is derived. {In addition, Trotterization with coarse steps leads to a recurrence of the initial state in the evolution;  the gap-ratio distribution becomes `un-chaotic' at even later times, after initially approaching a GUE distribution. This is shown in \Fig{fig:Trotter}, where we present numerically computed results, independent of the EHT-BW ansatz.}

{Figure~\ref{fig:BWgapratiostates} shows the gap-ratio averages separately for all six initial states. These are compared with experimental data (yellow circles) and with emulated data (blue squares). The results demonstrate that the emergence of a GUE regime is approximately similar for all initial states.}

In \Fig{fig:BWgapratiobases}, we study the accuracy of our tomography scheme concerning the number of random bases. Each basis is probed with $N_{\rm shots}=750$ samples, employing emulated data to eliminate the effect of device errors. Applying $N_\mathcal{U}=4$ to $N_\mathcal{U}=128$ random bases shows convergence in the lower-lying range of the entanglement spectrum relatively fast. However, the higher-lying part is not reconstructed even with a large sample size. The top panel displays the average EGRD over time, showing consistent behavior across the sample range. Error bars indicate the spread from combining initial states and symmetry sectors, which decrease with increasing $N_\mathcal{U}$. Concretely the error bars are obtained by computing the mean of the gap ratios for every sector and initial states separately, and then computing the standard deviation. 

Finally in \Fig{fig:BWgapratioshots}, we investigate the dependence on the number of measurement of probabilities $P_\mathcal{U}(s)$ in each basis (i.e., the number of shots), maintaining a constant number of measurement bases, $N_{U}=24$. Once more, convergence is evident in the lower segment of the ES in the lower panel, while the higher portion remains beyond reach even with infinite shots (red octagons). The top panel illustrates the EGRD, indicating consistency, within error bars, with the optimal BW-inspired ansatz result (cyan band).

{Each shot takes about 18 ms, implying a total raw measurement time of $5.5$ h for the data presented in Fig.~4 of the main text. Additional time is needed for calibration and reloading of ion chains following background-gas collisions. The actual Fig. 4 data was taken over six one-day runs. More shots mean that the random observables are more accurately determined. More random bases mean that more matrix elements of the state are effectively covered. With 32 physically allowed bitstrings, we simultaneously optimize 768 observables to generate the data presented in the main text. Not all are independent, however, because the randomly drawn bases are not necessarily orthogonal.}

In principle, given that the scheme is tomographically complete, we anticipate the ability to precisely reconstruct the entire ES with exponential resources. However in practice, we were not able to do so because of the significant bias of the cost function Eq.~(12) of the main text towards the low-energy portion of the entanglement spectrum. This bias places considerable strain on the numerical minimization routine, pushing it beyond its numerical-accuracy threshold. Despite this, its accuracy is satisfactory for analyzing experimental data, and we abstained from further optimization attempts. Exploring advanced optimization routines, including machine-learning techniques~\cite{huang2021demonstrating}, holds promise for EH tomography in future studies.

\end{document}